\documentclass[prd,showpacs,superscriptaddress,showkeys,floatfix,nofootinbib]{revtex4}
\usepackage{graphicx}
\usepackage{dcolumn}
\usepackage{bm}
\usepackage{color}
\usepackage{url}
\usepackage{amsmath,amssymb,graphicx}
\usepackage{amsmath}
\usepackage{amssymb}
\usepackage{mathrsfs}
\usepackage{accents}
\usepackage{epsfig}
\usepackage[lofdepth,lotdepth,caption=false]{subfig}
\usepackage[11pt]{moresize}
\usepackage{mathtools}
\usepackage{mathrsfs}
\newlength\figureheight 
\newlength\figurewidth 

\definecolor{MyGrey}{rgb}{0,0,0} 
\definecolor{MyDarkBlue}{rgb}{0.3,0.3,0.9} 
\definecolor{MyLightBlue}{rgb}{0.22,0.51,0.9}

\usepackage[colorlinks=true,linkcolor=MyDarkBlue,citecolor=MyDarkBlue,urlcolor=MyLightBlue,bookmarksnumbered=true,bookmarksopen]{hyperref}
\usepackage{breakurl} 
\usepackage[T1]{fontenc}
\newcommand{\ie}{\textit{i.e.,~}}
\newcommand{\eg}{\textit{e.g.,~}}

\def\beq{\begin{equation}}
\def\eeq{\end{equation}}

\begin{document}

\begin{flushright}
FERMILAB-PUB-17-150-T
\end{flushright}

\title{Visible Neutrino decay at DUNE}
\author{Pilar Coloma}
\email{pcoloma@fnal.gov}
\affiliation{Theory Department, Fermi National Accelerator Laboratory,
P.O. Box 500, Batavia, IL 60510, USA}

\author{O. L. G. Peres}
\email{orlando@ifi.unicamp.br}
\affiliation{Instituto de F\'isica Gleb Wataghin - UNICAMP, 13083-859, Campinas, SP, Brazil}
\affiliation{Abdus Salam International Centre for Theoretical Physics, ICTP, I-34010, Trieste, Italy}
\date{\today} 

\begin{abstract}
If the heaviest neutrino mass eigenstate is unstable, its decay modes could include lighter neutrino eigenstates. In this case part of the decay products could be visible, as they would interact at neutrino detectors via mixing. At neutrino oscillation experiments, a characteristic signature of such \emph{visible neutrino decay} would be an apparent excess of events at low energies. We focus on a simple phenomenological model in which the heaviest neutrino decays as $\nu_3 \rightarrow \nu_{1,2} + \phi$, where $\phi$ is a new light scalar. If neutrinos are Majorana particles the helicity-flipping decays would be observable (\ie $\nu \to \bar\nu + \phi$), leading to interesting observable consequences on the event rates. We compute the sensitivities of the Deep Underground Neutrino Experiment (DUNE) to the couplings of the new scalar as a function of the lightest neutrino mass. Under the assumption that only the heaviest neutrino is unstable, and for a normal mass ordering, we find that DUNE will be sensitive to values of $\tau_3/m_3 > 1.95 - 2.6\times 10^{-10}$~s/eV (90\% C.L.) (depending on the lightest neutrino mass), where $\tau_3$ and $m_3$ are the lifetime and mass of $\nu_3$, respectively. 
\end{abstract}

\pacs{14.60.Lm, 14.60.St, 14.60.Pq}
\keywords{neutrino decay, neutrino oscillation}
\maketitle

\section{Introduction}

Experimental neutrino physics has made great progress in the past two decades: we know have overwhelming evidence of neutrino oscillations and we know all mixing angles and mass splittings with precision, while at the same time the first hints for CP-violation in the neutrino sector are slowly emerging. Nevertheless, we still do not know what is the mechanism responsible of the generation of neutrino masses. A well-studied possibility is that they arise from Yukawa couplings with right-handed Majorana neutrinos and the Standard Model (SM) Higgs field. The right-handed Majorana masses could come from a new coupling to a scalar singlet (also known as the \emph{Majoron}) which takes a vacuum expectation value~\cite{Gelmini:1980re,Chikashige:1980ui}. In addition to generating neutrino masses, Majoron interactions can potentially lead to a number of interesting consequences, see \eg Refs.~\cite{Gelmini:1980re,Chikashige:1980qk,Chikashige:1980ui,Bertolini:1987kz,Santamaria:1987uq,Dias:2005jm}.  One of them is the possibility of neutrino decay, which is the focus of the present work: off-diagonal couplings in the form $g_{ij} \bar\nu_i \nu_j \phi$ could lead to the neutrino decay $\nu_i \to \nu_j + \phi$, where $i,j$ label the mass eigenstates. From the phenomenological point of view, we can distinguish between two different types of neutrino decay:
\begin{description}
\item[Invisible decay:] if the neutrino in the final state is not observable in the detector. This would be the case, for instance, if the neutrino in the final state is a light sterile neutrino (lighter than the active ones, in order to allow the decay kinematically). Another possibility is that the decay product involves an active neutrino which is however unobservable due to its low energy. Invisible neutrino decay would just lead to a depletion of the event rates at the detector with respect to the SM expectation.
\item[Visible decay:] if the neutrinos in the final state are active and detected. In this case, the decay products would affect the observable event rates at the detector, leading to a very different phenomenology with respect to the invisible decay scenario. Since the decay products will have lower energies than the parent particle, a typical signature of visible neutrino decay would be a pile-up of events at low energies.
\end{description}

Solar neutrinos have been used to place a bound on the lifetime of $\nu_2$~\cite{Berezhiani:1991vk,Acker:1992eh,Berezhiani:1992ry,Choubey:2000an,Beacom:2002cb,Joshipura:2002fb,Bandyopadhyay:2002qg,Berryman:2014qha,Picoreti:2015ika}. In most cases, these were derived under the assumption that the neutrino decays invisibly. Additional bounds~\cite{Frieman:1987as} have been obtained from the observation of supernova 1987A~\cite{Hirata:1987hu,Bionta:1987qt}. These mostly apply to the lifetimes of $\nu_1$ and $\nu_2$, as they are obtained from $\bar\nu_e$ observations.  Conversely, the lifetime of $\nu_3$ is much more difficult to constrain.  So far, bounds on the lifetime of $\nu_3$ have been derived from a fit to atmospheric and long-baseline oscillation data~\cite{GonzalezGarcia:2008ru}:
\[
\tau_3/m_3 > 2.9 \times 10^{-10}~\textrm{s}/\textrm{eV} \quad \textrm{at~90~\%~C.L.},
\]
where $\tau_3$ is the lifetime of $\nu_3$ in the center-of-mass frame, and $m_3$ is its mass. Additionally, in the decay scenario the neutrino propagation is not unitary, even in invisible mode, and then not only the charged current  interactions are modified but also the neutral current interactions. Recent analysis of the long-baseline experiments MINOS (using the full charged- and neutral-current data sample) and T2K (charged-current data) imply~\cite{Gomes:2014yua}
\[
\tau_3/m_3 > 2.8 \times 10^{-12}~\textrm{s}/\textrm{eV} \quad \textrm{at~90~\%~C.L.}. 
\]

In principle, the high-energy astrophysical neutrino sample from the Icecube experiment~\cite{Aartsen:2014gkd} could be used to set a bound on the lifetime of $\nu_3$ using the observed neutrino flavor ratios on Earth~\cite{Maltoni:2008jr,Baerwald:2012kc,Dorame:2013lka,Pagliaroli:2015rca,Shoemaker:2015qul,Bustamante:2015waa}. However, these are subject to large uncertainties coming \eg from the unknown initial flavor composition at the source, see e.g. Ref.~\cite{Bustamante:2016ciw} for a recent discussion. Finally, it was recently noted that the lifetime of $\nu_3$ can be constrained using future medium-baseline reactor experiments such as JUNO~\cite{He:2014zwa} or RENO-50~\cite{Kim:2014rfa}. In particular, the authors of Ref.~\cite{Abrahao:2015rba} found that the JUNO setup would be able to set the constraint $\tau_3/m_3 > 7.5\times 10^{-11}$~s/eV at 95\% C.L. .
 
In all cases mentioned above, the decay products were assumed to be invisible. Alternatively, the decay products could be visible due to their interactions with the SM fermions via mixing. For example, if the decay $\nu_3 \to \nu_2 + \phi $ is allowed, a $\nu_2$ mass eigenstate ($\nu_2=U_{e2}^*\nu_e+U_{\mu2}^*\nu_{\mu}+U_{\tau2}^*\nu_{\tau}$)  would have a non-vanishing cross section to produce an electron at neutrino detectors thanks to the $U_{e2}$ element in the mixing matrix~\cite{Lindner:2001fx}. This may render the decay products observable at neutrino experiments and would lead to different signatures than those expected in the invisible decay scenario. For example, in Ref.~\cite{PalomaresRuiz:2005vf}, the authors considered the possibility that a heavy sterile neutrino (produced in pion decays via small mixing with muon neutrinos) could rapidly decay to an active neutrino plus a Majoron. They found that the excess of events observed at LSND experiment may be explained by the interactions of the active decay products at the detector. Moreover, if neutrinos are Dirac particles Majoron models would allow for decays with a change in helicity, such as $\nu_i \to \bar\nu_j + \phi$. These signatures would be easy to distinguish at magnetized detectors, but they would be more challenging to distinguish at liquid Argon detectors such as the DUNE~\cite{Acciarri:2016crz} or the MicroBooNE detectors~\cite{Acciarri:2016smi}. Another relevant imprint of neutrino decay would be a shift in the energy profile of the beam, leading to an apparent excess of events at low energies. The observable effect would be larger at on-axis neutrino beam experiments, since the observable event rates at low energies will receive contributions from the parent neutrinos initially produced at higher energies. Consequently, the DUNE detector, with its very low expected detection threshold and wide-band neutrino beam, would be very well-suited to search for this kind of signature. Also, for experiments that cross the earth matter, significant changes in the oscillation behaviour are expected due to the Mikheev-Smirnov-Wolfenstein (MSW) effect~\cite{Mikheev:1986wj,Wolfenstein:1977ue}. The decay scenario in presence of a matter potential has been discussed in the literature in the context of solar and supernovae neutrinos~\cite{Berezhiani:1989za,Berezhiani:1991vk,Giunti:1992sy,Berezhiani:1993iy,1402-4896-64-3-004}, but not for long-baseline experiments. 

The paper is structured as follows. In Sec.~\ref{sec:framework} we describe the general framework used and the expected experimental signatures. Section~\ref{sec:exp} summarizes the simulation details and the effects of the decay on the expected event rates. In Sec.~\ref{sec:results} we derive the expected sensitivity of DUNE to visible neutrino decay and compare to other limits, and we conclude in Sec.~\ref{sec:conclusions}. Finally, some technical details regarding the computation of the oscillation probabilities in presence of visible neutrino decay are summarized in the Appendix.

\section{Neutrino decay phenomenology}
\label{sec:framework}

For concreteness, we will assume that the neutrino interaction with a Majoron $\phi$ is responsible of neutrino decay. In order to be as general as possible, we will allow both scalar and pseudo-scalar couplings. The Lagrangian responsible for the new interactions reads
\beq
\label{eq:lagrangian}
{\cal L}_{\rm int}=\sum_{i=1,2}\dfrac{g_{3i}}{2}\bar{\nu}_i \nu_3 \phi +\dfrac{g_{3i}^{\prime}}{2}\bar{\nu}_i i\gamma_5 \nu_3 \phi + \textrm{h.c.}\, ,
\eeq
where $g_{3i} (g_{3i}^{\prime})$ are the scalar (pseudo-scalar) couplings, which in principle can be complex. 
The couplings introduced in Eq.~\eqref{eq:lagrangian} would lead to neutrino decay in the form $\nu' \to \nu + \phi$. It is worth stressing out that in the case of Majorana neutrinos, neutrino decays with same helicity ($ \nu \rightarrow \nu+ \phi$, or $\bar\nu \rightarrow \bar \nu+ \phi$) or with \emph{opposite} helicity ($\nu \rightarrow \bar\nu +\phi$ or $\bar\nu \rightarrow \nu+ \phi$) can be observed. Conversely, if neutrinos are Dirac, the neutrinos produced with opposite helicity will not interact in the detector and, thus, this decay mode would be unobservable. The decay modes with opposite helicity can have important phenomenological consequences, as we will discuss later on.  For concreteness, throughout this work we will assume that the only neutrino that is unstable is $\nu_3$, a normal mass ordering scheme ($m_3 > m_2 > m_1$), and a massless Majoron.

\subsection{Decay widths}
\label{sec:widths}
 
\begin{figure}
\begin{center}
\includegraphics[scale=0.65]{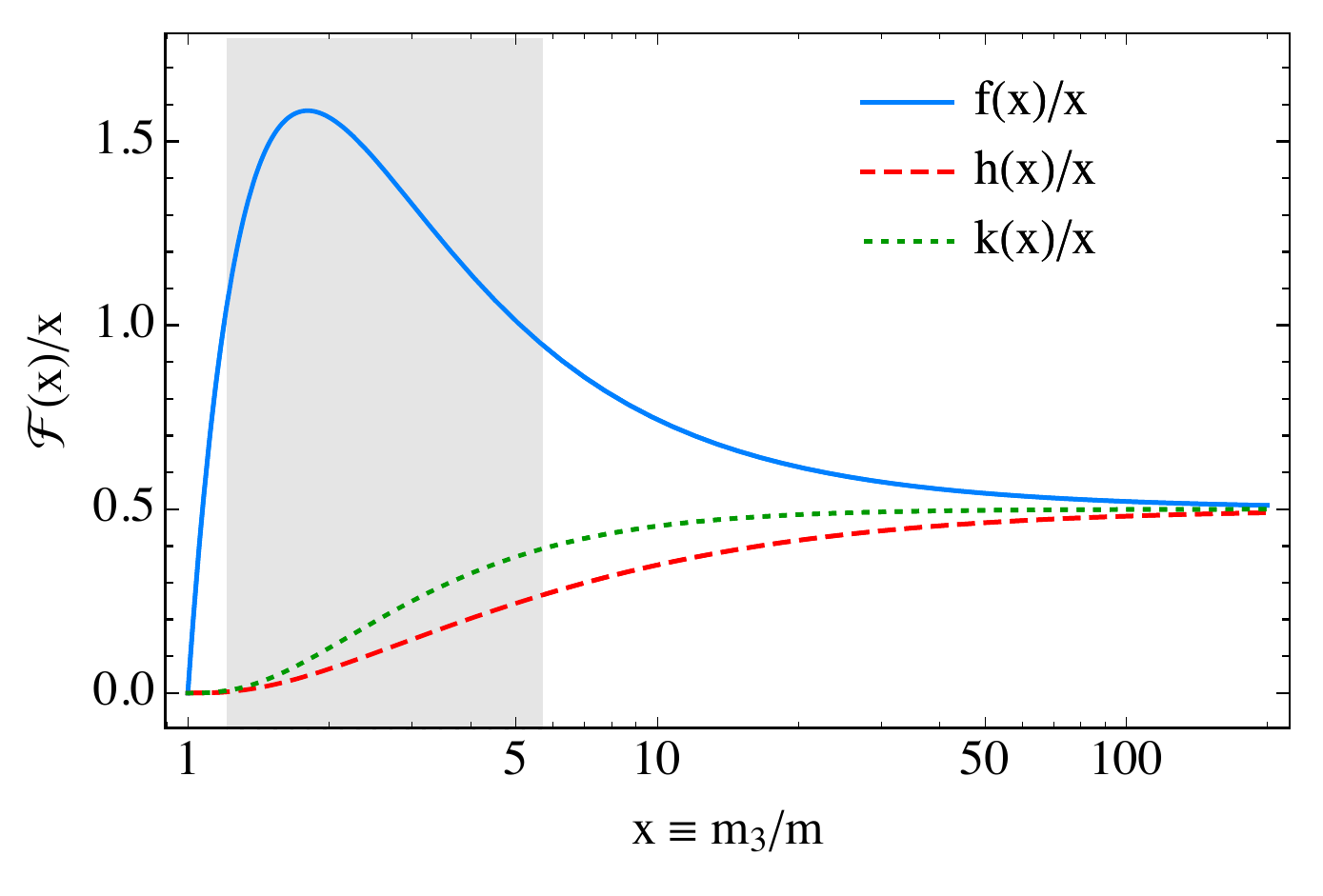}
\caption{\label{fig:fgk} Dependence of the functions $f$, $h$ and $k$ defined in Eq.~\eqref{eq:fgk} with the value of $x_i\equiv m_3/m_i$, for the decay $\nu_3 \rightarrow \nu_i$. The branching ratios for the helicity-conserving decays directly depend on the functions $f$ and $h$, while the value of $k$ determines the branching ratio for the helicity-flipping decay, see Eq.~\eqref{eq:channels}. The shaded area indicates the region allowed for $x_2$, after taking into account the values of the mass splittings determined from oscillations as well as the upper limit from PLANCK + BOSS BAO on the sum of the active neutrino masses, $\sum_i m_i < 0.23$~eV~\cite{Ade:2013zuv}. For $\nu_1$ there is no upper limit on the value of $x_1$; the lower limit is very similar to the one for $x_2$ and is therefore not shown.
}
\end{center}
\end{figure}

From the Lagrangian in Eq.~\eqref{eq:lagrangian}, the neutrino decay widths in the laboratory  system for the different decay modes can be computed as~\cite{Kim:1990km}: 
\begin{eqnarray}
\Gamma^{\rm S}(\nu_3\to\sum_{i=1,2}\nu_i) & = & 
\dfrac{m_3^2}{16\pi E} \sum_{i=1,2}g_{3i}^2 \left(\dfrac{f(x_i)}{x_i} \right)\Theta(x_i-1)\, , \nonumber \\
\Gamma^{\rm PS}(\nu_3\to \sum_{i=1,2}\nu_i)) & = & \dfrac{m_3^2}{16\pi E} \sum_{i=1,2} (g^{\prime}_{3i})^2 \left(\dfrac{h(x_i)}{x_i}\right)\Theta(x_i-1) \, , \label{eq:channels} \\
\Gamma^{\rm S+PS}(\nu_3\to  \sum_{i=1,2}\bar{\nu}_i)) & = & \dfrac{m_3^2}{16\pi E}  \sum_{i=1,2} \left[g_{3i}^2+(g^{\prime}_{3i})^2\right]\left(\dfrac{k(x_i)}{x_i}\right)\Theta(x_i-1) \nonumber \, . 
\end{eqnarray}
Here, $\Gamma^{\rm S}$ and $\Gamma^{\rm PS}$ stand for the decay widths obtained from the scalar and pseudo-scalar interactions in the Lagrangian in Eq.~\eqref{eq:lagrangian}. We have defined $x_i \equiv m_3/m_i$ as the ratio between the initial and final neutrino masses (with $i=1,2$), while $E$ is the energy of the initial neutrino ($\nu_3$). The $\Theta$ function enforces the kinematical limit where the decay is allowed, $m_3 > m_i$. The first two decay widths listed in Eq.~\eqref{eq:channels} correspond to the helicity-conserving decay, while the last one corresponds to the helicity-flipping mode. The functions $f(x)$, $h(x)$ and $k(x)$ 
are defined as~\cite{Kim:1990km}: 
\begin{eqnarray}
f(x) & = & \dfrac{x}{2}+2+\dfrac{2 \ln(x)}{x}-\dfrac{2}{x^2}-\dfrac{1}{(2x^3)} \, , \nonumber \\
h(x) & = & \dfrac{x}{2}-2+\dfrac{2 \ln(x)}{x}+\dfrac{2}{x^2}-\dfrac{1}{(2x^3)} \, , \label{eq:fgk}  \\
k(x) & = &\dfrac{x}{2}-\dfrac{2\ln(x)}{x}-\dfrac{1}{(2x^3)} \,. \nonumber
\end{eqnarray}
Figure~\ref{fig:fgk} shows the values of $f(x_i)$, $h(x_i)$ and $k(x_i)$ as a function of $x_i$. Two interesting limits are easily identified from the results shown in Fig.~\ref{fig:fgk}: (i) for $m_3 \to  m_i $ (\ie $x_i \to 1$) the scalar decay with same helicity dominates, since $f(x_i)\gg h(x_i), k(x_i)$ (see Eq.~\eqref{eq:channels}); and (ii) for $m_3 \gg  m_i $ (\ie $ x_i \to \infty$) all branching ratios in Eq.~\eqref{eq:channels} take similar values, as $f(x_i)\sim h(x_i)\sim k(x_i) $.  

The values of $x_i$ for the light neutrino masses are partially constrained since the two mass-squared splittings are precisely determined from neutrino oscillation data. Under the assumption of a normal ordering ($m_3 > m_2 > m_1$), the masses of the two heaviest neutrinos can be expressed as a function of the lightest neutrino mass as: 
\begin{align}
m_3^2 & = m_1^2 +\Delta m^2_{31} \, , \\
m_2^2 & = m_1^2 +\Delta m^2_{21} \, .
\label{mass}
\end{align} 
where $\Delta m^2_{ij} \equiv m_i^2-m_j^2$. From global fits to oscillation data~\cite{Bergstrom:2015rba,Esteban:2016qun} we have $\Delta m^2_{21} \sim 7.5 \times 10^{-5}$~eV$^2$ and $\Delta m^2_{31} \sim 2.5 \times 10^{-3}$~eV$^2$. Thus, for the decay $\nu_3\to \nu_2+ \phi$ the ratio $x_2 \equiv m_3 / m_2 $ can only take values in the range  $1 \lesssim x_2 \lesssim  6$, which is indicated by the shaded area in Fig.~\ref{fig:fgk}. The upper limit for $x_2$ is obtained using the best-fit values for the squared-mass differences and setting $m_1 = 0$. The lower limit corresponds to the case of quasi-degenerate neutrinos ($m_1 \simeq m_2 \simeq m_3$). It is limited not to be exactly $1$ by the contraints from cosmology, which impose an upper bound on the sum of neutrino masses $\sum_i m_i < 0.23$~eV~\cite{Ade:2013zuv}. Conversely, the branching ratios for the decay  $\nu_3\to \nu_1+ \phi$ depend on $x_1\equiv  m_3/m_1 $ which can take values in the range $1\lesssim x_1 < \infty$.

Following the arguments outlined above, it is easy to see that the phenomenology is essentially fixed by the scale of the lightest neutrino mass as this determines the values of $x_i$ and, consequently, the decay widths in Eq.~\eqref{eq:channels}. We can distinguish two well-defined limiting cases:
\begin{description}
\item[Hierarchical scenario ($m_1 \to 0$):] In this limit $x_1 \to \infty$ and the three kinematic functions for the decay $3 \to 1$ are approximately equal, see Fig.~\ref{fig:fgk}. Thus, the branching ratios for the helicity-flipping ($\nu_3\to \bar\nu_1 +\phi$) and helicity-conserving ($\nu_3\to \nu_1 +\phi$) decays will all be comparable as long as the scalar ($g$) and pseudo-scalar ($g'$) couplings are equal. However in this scenario the value of $x_2$ is bounded from above as $x_2 \to 6$. Thus, the decays $3\to 2$ will be dominated by the helicity-conserving width, see Fig.~\ref{fig:fgk}, with a larger contribution from the decay produced via the scalar coupling (solid blue line), the pseudo-scalar branching ratio being much smaller (dashed red line).
\item[Quasi-degenerate scenario ($m_1 \to m_2,m_3$):] In this limit both $x_{1,2} \to 1$. As can be seen from the comparison between the different lines in Fig.~\ref{fig:fgk}, the decay will be mainly dominated by the helicity-conserving decay which takes place via the scalar coupling (solid blue curve), while the other two branching ratios will be largely suppressed.
\end{description}
From this simplified description it is easy to see that, in general, we expect that the sensitivity to the decay will mainly come from the helicity conserving decay with scalar coupling, except in the limit of vanishing $m_1$ when the helicity-flipping decays will turn on and contribute to the sensitivity. This will be confirmed by the results of our numerical simulations in Sec.~\ref{sec:results}. In between these two regimes, the phenomenology will generally be more complex, as the functions $f$, $h$ and $k$ depend on $x_1$ and $x_2$, which in turn depend on $m_1$, $\Delta m^2_{21}$ and $\Delta m^2_{31}$. Finally, it should be noted that in case the scalar and pseudo-scalar couplings are very different, the situation can be more complex and the phenomenology might differ from the cases outlined above. 

\begin{figure}
\begin{center}
\includegraphics[scale=0.6]{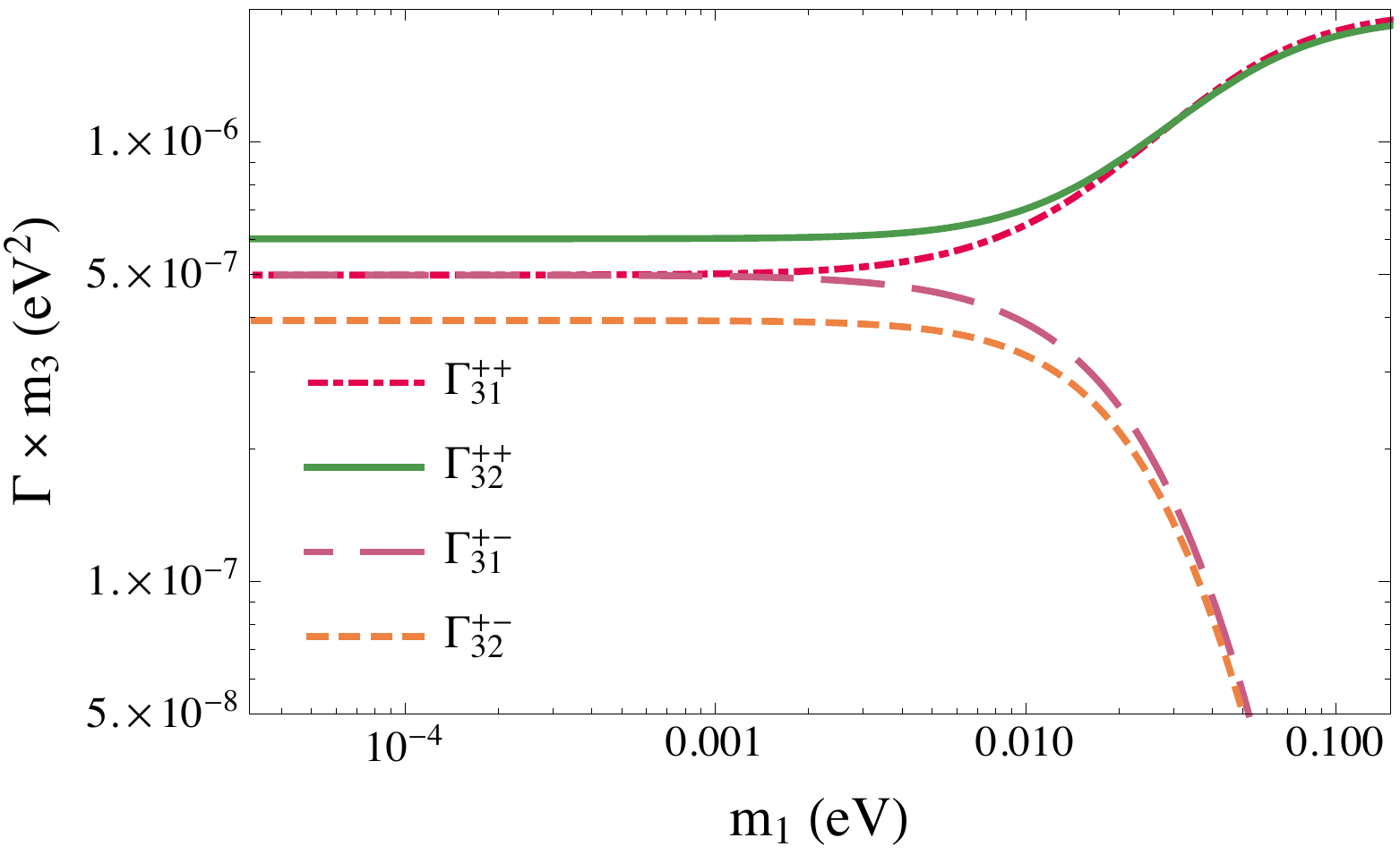}
\caption{\label{fig:gammas}Dependence of the product of the decay width times the mass of $\nu_3$ with the mass of the lightest neutrino, for the different decay modes, computed in the center-of-mass frame. The assumed values of the couplings are $|g_{3i}| = |g_{3i}'| = 0.1$, for $i=1,2$. In the labels, $ \Gamma_{3j}^{rs} \equiv \Gamma (\nu_3^r \to \nu_j^s)$, where $rs =++$ and $rs=+-$ indicate if the helicity is conserved or flipped in the decay. Neutrino oscillation constraints on the mass-squared splittings have been accounted for, see Eq.~\eqref{mass}. 
}
\end{center}
\end{figure}

For illustration, the behaviour of the decay width with the mass of the lightest neutrino is shown in Fig.~\ref{fig:gammas}, for the different decay modes of the neutrino, $ \Gamma_{3j}^{rs} \equiv \Gamma (\nu_3^r \to \nu_j^s)$, where $rs =++$ and $rs=+-$ indicate if the helicity is conserved or flipped in the decay(in our notation, $ r=+/-$ indicates neutrino/antineutrino).  In fact, we show the product  $\Gamma_3 \times m_3$ since this relates to the decay width in the lab frame, which is what neutrino oscillation experiments can actually probe. This figure will help to explain the dependence of the sensitivity to neutrino decay with the mass of $m_1$, as we will discuss in Sec.~\ref{sec:results}. 

\subsection{Invisible decay}
\label{sec:invisible}

Let us first consider the case where the decay products are unobservable at the detector, either because their energy is too low or because they are sterile. This leads to the so-called \emph{invisible neutrino decay}, as outlined in the introduction. The effect of invisible decay on the neutrino oscillation probabilities has been discussed previously in the literature, see for instance Refs.~\cite{Frieman:1987as,Barger:1998xk,Fogli:1999qt,Lindner:2001fx,Beacom:2002cb,Joshipura:2002fb,Bandyopadhyay:2002qg,Ando:2003ie,Ando:2004qe,Fogli:2004gy,PalomaresRuiz:2005vf,Maltoni:2008jr,GonzalezGarcia:2008ru,Baerwald:2012kc,Dorame:2013lka,Gomes:2014yua,Berryman:2014qha,Picoreti:2015ika,Abrahao:2015rba,Bustamante:2015waa,Bustamante:2016ciw}. Its main impact consists on a depletion of the number of events with respect to the expectation in the standard scenario. The effect in this case will only depend on the total decay width, computed as the sum of the different contributions listed in Eq.~\eqref{eq:channels}.

\subsubsection{Vacuum case}
The Hamiltonian in vacuum takes the form:
\begin{equation}
H = U^{r} \left( \begin{array}{ccc} 
0 & 0 & 0 \\
0 & \tfrac{\Delta m^2_{21}}{2E} & 0 \\
0 & 0 & \tfrac{\Delta m^2_{31}}{2E} - i\Gamma_3 /2
\end{array} \right)  (U^{r})^\dagger
\label{eq:Hvac}
\end{equation}
where we have used the relativistic approximation $E_i \simeq p + \dfrac{m^2_{i}}{2E}$ and we have subtracted an overall constant to the system. Here, $U^r$ stands for the leptonic mixing matrix in its usual form, $r$ indicates the helicity, and $\Gamma_3 \equiv  1/\tau_3^{\rm lab} =m_3/(\tau_3 E )$ is the \emph{total} decay width of $\nu_3$  in the laboratory frame, where $\tau_3$ is the neutrino lifetime in  its rest frame. The index $r$ is used to distinguish neutrinos from antineutrinos ($r =  +/-$ for neutrinos/antineutrinos), which will be important later on for the discussion of matter effects.

The oscillation probabilities affected by the invisible decay of $\nu_3$ can be easily computed as in the standard case, replacing $\frac{\Delta m^2_{31}L}{2E} \to  \frac{\Delta m^2_{31}L}{2E} -i \frac{\Gamma_3 L}{2} $. For the appearance channels ($\alpha \neq \beta$), the probabilities in vacuum read as~\cite{Abrahao:2015rba,Lindner:2001fx}:
\begin{eqnarray}
P^{\rm inv}_{ \nu_\alpha^r \to \nu_\beta^r } (E) & = & \bigg | U^{r}_{\alpha 1}(U^r_{\beta 1})^* + U^r_{\alpha 2}(U^{r}_{\beta 2})^*e^{-i\Delta m^2_{21}L/(2E)}  + U^r_{\alpha 3}(U^{r}_{\beta 3})^* e^{- i\Delta m^2_{31}L/(2E)} e^{-\Gamma_3 L/2} \bigg |^2 =  \nonumber \\
& = & P(\nu_\alpha^r \to \nu_\beta^r ; \textrm{no\, decay}) - | U_{\alpha 3}^{r}|^2 | U_{\beta 3}^{r}|^2 (1 - e^{-\Gamma_3 L }) 
-2 \sum_{i = 1,2} \textrm{Re}[\mathcal{C}^{r}_{\alpha\beta i}]\cos \left(\frac{\Delta m^2_{3i} L}{2E} \right) \left( 1 - e^{-\Gamma_3 L/2}\right) + \nonumber \\ & &
+2 \sum_{i = 1,2} \textrm{Im}[\mathcal{C}^{r}_{\alpha\beta i}]\sin \left( \frac{\Delta m^2_{3i} L}{2E}\right) \left( 1 - e^{-\Gamma_3 L/2}\right) \, , 
\label{eq:PinvVac}
\end{eqnarray}
where $\mathcal{C}^{r}_{\alpha\beta i } \equiv U^r_{\alpha i}(U_{\beta i}^r)^*(U_{\alpha 3}^r)^*U^r_{\beta 3} $. From Eq.~\eqref{eq:PinvVac} it can be seen that the invisible decay generally leads to a depletion in the number of events. In the case where the solar mass-squared difference is neglected it can be shown that for the $\nu_\mu \to \nu_e$ appearance channel the effect is proportional to $|U_{\mu 3}|^2|U_{e 3}|^2 \sim s_{23}^2 c_{13}^2 s_{13}^2 \sim \mathcal{O}(10^{-2})$, see Ref.~\cite{Abrahao:2015rba}. This coefficient is of the same size as the leading order term in the oscillation probability for this channel. As can be seen from Eq.~\eqref{eq:PinvVac}, the effect of invisible decay is reduced to the impact of one effective parameter on the probabilities, $\tau_3 / m_3$. 

\subsubsection{Matter effects}

For long-baseline experiments like DUNE~\cite{Acciarri:2016crz} it is essential to include matter effects in the formalism, as they significantly affect the neutrino oscillation probabilities. In the case of invisible decay, this means that the Hamiltonian of the system will now read as:
\begin{equation}
H = U \left( \begin{array}{ccc} 
0 & 0 & 0 \\
0 & \tfrac{\Delta m^2_{21}}{2E} & 0 \\
0 & 0 & \tfrac{\Delta m^2_{31}}{2E} - i\frac{\Gamma_3 }{2}
\end{array} \right) U^\dagger + A\left( 
\begin{array}{ccc} 
1 & 0 & 0 \\
0 & 0 & 0 \\
0 & 0 & 0 \\
\end{array} 
\right) \, ,
\label{eq:Hmat}
\end{equation}
where the second term includes the matter potential $A$ felt by the neutrinos as they propagate in a medium of constant density. Here, $A\equiv \pm\sqrt{2}G_F N_e$, where $G_F$ is the Fermi constant and $N_e$ is the number density of electrons in the medium. The  $+/- $ sign corresponds to neutrinos/antineutrinos.
  
In the limit of a stable neutrino, the Hamiltonian in Eq.~\eqref{eq:Hmat} can be diagonalized with a unitary matrix in  matter, $\tilde{U}^{r}$, which will generally differ from the one that diagonalizes the system in vacuum. The eigenvalues of the Hamiltonian are also affected by the matter potential, and are denoted as $\Delta \tilde{m}^2_{ij}$. In what follows, our notation will explicitly reflect the presence of a matter potential: for a given quantity/observable X in vaccum, we will denote its corresponding value in a system in presence of a matter potential by $\tilde{X}$.

In the case of an unstable neutrino, the Hamiltonian in  Eq.~\eqref{eq:Hmat} is no longer Hermitian and it cannot be diagonalized by a unitary transformation. This implies that  $(\tilde{U}^{r})^{-1} \neq (\tilde{U}^{r} )^{\dagger} $ in the general case. The oscillation probabilities are obtained after diagonalization of the Hamiltonian in Eq.~\eqref{eq:Hmat}. Taking this into consideration, the oscillation probabilities in presence of invisible decay are written in the matter regime as:
\begin{equation}
\tilde{P}^{\rm inv}_{\nu_\alpha^r \to \nu_\beta^r} (E) = \bigg |
\tilde{U}^{r}_{\beta 1}(\tilde{U}^{r})^{-1}_{1 \alpha}  + 
\tilde{U}^{r}_{\beta 2}(\tilde{U}^{r})^{-1}_{2 \alpha} e^{-i\Delta \tilde{m}^2_{21} L / (2E)} + 
\tilde{U}^{r}_{\beta 3}(\tilde{U}^{r})^{-1}_{3 \alpha} e^{-i\Delta \tilde{m}^2_{31} L / (2E)} e^{-\tilde{\Gamma}_3 L/2} \bigg |^2 , 
\label{eq:PinvMatter}
\end{equation}
where it should be kept in mind that the width $\Gamma_3$ also depends on the neutrino energy $E$. An important difference with respect to the case of neutrino decay in vacuum is that the neutrino width in this case is {\em also modified} by the presence of the matter potential. This is the case because the width depends on the mass of the third eigenstate, which in presence of matter effects is modified as 
\[
m_3 = \sqrt{ m_1^2 + \Delta m^2_{31}}   \longrightarrow \tilde{m}_3 = \sqrt{ m_1^2 +\Delta  \tilde{m}^2_{31}} \, ,
\] 
affecting the total width as
\[
\Gamma_3  \longrightarrow  \tilde{\Gamma}_3\equiv \tilde{m}_3/(\tau_3 E)  \, .
\] 
Moreover we note that, in matter, the total width will generally be different for neutrinos and antineutrinos as they are affected differently by the potential.

\subsection{Visible decay}
\label{sec:visible}

If the decay products are active, the resulting neutrino could in principle interact at the detector, giving a \emph{visible} signal. The final observability of such effect will eventually depend on the energy of the final decay products as well as on the detection threshold at low energies and other experimental factors. The effect of visible neutrino decay on the observable event rates has been studied previously in the literature, see \eg Refs.~\cite{Berezhiani:1991vk,Lindner:2001fx,PalomaresRuiz:2005vf}. In the remainder of this section, we will follow the formalism outlined in Refs.~\cite{Lindner:2001fx} and~\cite{PalomaresRuiz:2005vf}, paying special attention to the impact of the matter potential on the resulting probability. We will also explicitly consider the impact of the decay on the energy distribution of the final neutrino observed at the detector.

\subsubsection{Vacuum case}

The differential probability that a neutrino produced in a flavor state $\nu_\alpha$ with initial energy $E_\alpha$ is detected as a neutrino with flavor $\beta$ and final energy $E_\beta$ due to the combined effect of oscillations plus the decay $\nu_3 \rightarrow \nu_{1,2} + \phi $ can be expressed as~\cite{PalomaresRuiz:2005vf}:
\begin{eqnarray}
\frac{d\mathcal{P}_{\nu_\alpha^r \to \nu_\beta^s}}{dE_\beta} & = & 
P_{\nu_\alpha^r \to \nu_\beta^s}^{\rm inv}(E_\alpha)\, \delta(E_\alpha - E_\beta) \delta_{rs}
 +  \Delta P^{\rm vis}_{ \nu_\alpha^r \to \nu_\beta^s } (E_\alpha, E_\beta) \, .
\label{eq:P1}
\end{eqnarray}
Here the first term corresponds to the standard probability modified by the decay as in Eq.~\eqref{eq:PinvVac} and only takes place when the initial and final helicities coincide. Conversely, the second term includes the visible contribution: it is only present if the decay products have interactions with the SM fermions. The helicities of the initial ($r$) and final ($s$) neutrinos can in principle be different in this case. 

Let us consider a decay taking place at a distance $L' < L$ from the neutrino source.\footnote{In the most general case, the decay will lead to a certain opening angle for the final decay products with respect to the direction of the initial neutrino. As a consequence the final neutrino may not enter the detector. This will ultimately depend on the detector size, its distance to the source and the initial neutrino energy. However, in the case of long-baseline neutrino experiments the detector acceptance for the decay products is always extremely good~\cite{Lindner:2001fx}. } Neglecting matter effects, the visible contribution from this process to the probability can be easily computed starting from the following amplitude~\cite{Lindner:2001fx}: 
\begin{equation}
\mathcal A_{\nu_\alpha^r \to \nu_\beta^s} (E_\alpha, E_\beta) = \sum_{j= 1,2} (U^r_{\alpha 3})^* U^s_{\beta j}
e^{-i E_\beta (L - L')} e^{-i E_\alpha L'} e^{-\Gamma^{rs}_{3j} L'/2}\sqrt{\Gamma^{rs}_{3j}}\sqrt{W^{rs}_{3j}}\, .
\label{eq:Avac}
\end{equation}
Here, $W^{rs}_{3j} \equiv \tfrac{1}{\Gamma_{3j}^{rs}}\tfrac{d\Gamma^{rs}_{3j} (E_\alpha, E_\beta)}{dE_{\beta}}$ is the normalized energy distribution for the decay $\nu_3^r \to \nu^s_j +\phi$ (see App.~\ref{app:decay} for details) and $\alpha$ ($\beta$) denotes the flavor of the initial (final) neutrino, which has energy $E_\alpha$ ($E_\beta$). Finally, $\Gamma^{rs}_{3j} $ is the partial width for the decay $\nu_3^r \rightarrow \nu_j^s + \phi$ in Eq.~\eqref{eq:channels} and therefore only depends on the initial neutrino energy $E_\alpha$. 

A comment regarding the notation used in Eq.~\eqref{eq:Avac} is needed at this point. Even though we use flavor indices to refer to the energies in the initial and final states, these are quantities which are only well-defined in the mass basis. Strictly speaking, $E_3$ should be used instead of $E_\alpha$, and $E_{1}$ or $E_2$ should be used instead of $E_\beta$, depending on the mass eigenstate produced in the decay. Our notation in terms of flavor indices is chosen to stress the fact that the detection process takes place in the flavor basis and, thus, both $\nu_1$ and $\nu_2$ can contribute to the final observable quantity (the number of charged-current events where a charged lepton with flavor $\beta$ is observed).

The total contribution to the differential probability in Eq.~\eqref{eq:P1} coming from the visible decay is obtained after integration over all possible values of $L'$ between the source and the detector, as:
\begin{align}
\Delta P^{\rm vis}_{\nu_\alpha^r \to \nu_\beta^s } (E_\alpha, E_\beta) & = \int_0^L \mid \mathcal A_{\nu_\alpha^r \to \nu_\beta^s} (E_\alpha, E_\beta)\mid^2 dL'   = \nonumber \\
& = \sum_{j = 1,2} \big | (U^r_{\alpha 3})^* U^s_{\beta j} \big |^2 \, W^{rs}_{3j} \, \left( 1 - e^{-\Gamma^{rs}_{3j} L}\right) + \nonumber \\
& \phantom{laa} +
4 \big | U^r_{\alpha 3} \big |^2 \big |  U^s_{\beta 1}(U^s_{\beta 2})^* \sqrt{(\Gamma^{rs}_{31})^*\Gamma^{rs}_{32}} \big | \, \sqrt{W^{rs}_{31}W^{rs}_{32}}   \,   \,
 \frac{ E_\beta}{ E_\beta^2 (\Gamma^{rs}_{31} +  \Gamma^{rs}_{32})^2 + (\Delta m^2_{21})^2} 
\times  \nonumber \\ 
& \phantom{laa}\times \left\{  \Delta m^2_{21} 
\left[ \sin\left( \xi + \frac{\Delta m^2_{21}L}{2 E_\beta}\right) - \sin\xi \, e^{-(\Gamma^{rs}_{31} + \Gamma^{rs}_{32}) L/2}\right]  
\right. + \nonumber \\ 
& \phantom{laaaa} \left. + \;
  E_\beta (\Gamma^{rs}_{31} + \Gamma^{rs}_{32}) \left[  \cos\left( \xi + \frac{\Delta m^2_{21}L}{2 E_\beta}\right) - \cos \xi \, e^{-(\Gamma^{rs}_{31} + \Gamma^{rs}_{32}) L/2} \right]   \right\} \, . \label{eq:Pvis} 
\end{align}
Here we have defined the complex phase $\xi \equiv \textrm{arg}\left(U_{\beta 1}^s (U_{\beta 2}^s)^* \sqrt{(\Gamma^{rs}_{31})^*\Gamma^{rs}_{32}}\right) $. The first line comes from the contribution given by $\nu_3 \to \nu_i + \phi$, while the rest of the expression contains the interference between the two amplitude contributions from $\nu_3\to \nu_2 + \phi $ and $\nu_3\to \nu_1 + \phi $. The interference will only take place in the region of parameter space where both decays are kinematically possible. For simplicity in our simulations we will assume that the CP-violating phase $\xi$ vanishes. 

\subsubsection{Matter regime}

In the matter regime the visible decay contribution can be derived in the same way as the vacuum case, starting from the following amplitude instead: 
\begin{equation}
\widetilde{\mathcal{A}}_{\nu_\alpha^r \to \nu_\beta^s}(E_\alpha, E_\beta) = \sum_{j= 1,2} 
(\widetilde{U}^{r})^{-1}_{3 \alpha} \tilde{U}^s_{\beta j}
e^{-i E_\beta (L - L')} e^{-i E_\alpha L'} e^{-\tilde{\Gamma}^{rs}_{3j}L'/2}\sqrt{\tilde{\Gamma}^{rs}_{3j}}
\sqrt{\tilde{W}^{rs}_{3j}} \, ,
\label{eq:Amatter}
\end{equation}
which is obtained from Eq.~\eqref{eq:Avac} just replacing the decay widths and eigenvalues by the corresponding ones in matter ($U^{r}\to \tilde{U}^{r}$, $\Delta m^2_{ij} \to \Delta \tilde{m}^2_{ij}$, $\Gamma_3 \to \tilde{\Gamma}_3$).  In this case the diagonalization is no longer done via a unitary matrix and, therefore, $(\tilde{U}^{r})^{-1} \neq (\tilde{U}^{r})^{\dagger}$. It should be noted that both $\tilde{U}$ and $\Delta \tilde{m}^2_{ij}$ will now ultimately depend on the neutrino energy, as they depend on the matter potential. 

The computation of the contribution to the differential probability be done following the same steps as in the vacuum case (with some subtleties), see App.~\ref{app:decay} for details.  However, several differences among the vacuum and matter cases should be noted at this point. First, the matrices which diagonalize the system for neutrinos and antineutrinos will now be generally different, as they are affected by a different matter potential. This is especially relevant when computing the visible contribution due to helicity-flipping decays. Another important different with respect to the vacuum case is that, since the masses $m_3$ and $m_2$ will change in presence of a matter potential, the regions where the decay is kinematically allowed will also depend on $A$. This will be relevant for the computation of the number of events, and is discussed in more detail in Sec.~\ref{sec:exp}.

\section{Simulation details and expected event rates}
\label{sec:exp}

Once the effect of neutrino decay is considered, the events observed in a given energy bin $\Delta E_{\beta,i}$ will receive contributions from neutrinos at higher energies which have decayed emitting a Majoron. Therefore, the final observable event distribution should be computed after integration over all values of $E_\alpha$ above $E_\beta$, turning into a pile-up of events at low energies. Therefore, one of the key factors will be the low energy thresholds. Moreover, if decay modes with opposite helicity are allowed, they could lead to an observed excess of \emph{anti-neutrino} events at low energies in the neutrino running mode, or viceversa. At a magnetized detector, the opposite helicity contributions to the total event rates would be easily distinguished. However, at DUNE~\cite{Acciarri:2016crz} the detector is not magnetized, so the event rates will receive contributions from the decay products with equal and  opposite helicity\footnote{Some charge identification may be feasible using statistical discriminators. However, for simplicity we ignore these in this work. }.The resulting expression for the differential number of events for DUNE, as a function of the neutrino energy in the final state $E_\beta$, reads as
\begin{eqnarray}
\frac{dN^s}{dE_\beta} & = & \sigma^s_\beta(E_\beta) \phi^s_\alpha (E_\beta) 
\tilde{P}^{\rm inv}_{\nu_\alpha^s \to \nu_\beta^s} (E_\beta) +  
\sigma^s_\beta(E_\beta) \int_{E_\beta}^{\tilde{E}_{\rm max}} dE_\alpha \phi^s_\alpha (E_\alpha) \Delta \tilde{P}_{\nu_\alpha^s \rightarrow \nu_\beta^s }^{\rm vis}(E_\alpha,E_\beta) + \nonumber \\
& + & 
\sigma^r_\beta(E_\beta) \int_{E_\beta}^{\tilde{E}_{\rm max}} dE_\alpha \phi^s_\alpha (E_\alpha) \Delta \tilde{P}_{\nu_\alpha^s \rightarrow \nu_\beta^r}^{\rm vis}(E_\alpha,E_\beta) \,,
 \label{eq:dNdE}
\end{eqnarray}
and includes the impact of oscillations and decay. The upper limits of the integrals in the second and third terms are computed from two body decay kinematics taking into account the masses of the initial and final neutrino eigenstates and assuming a massless Majoron, $\tilde{E}_{\rm max}=\tilde{x}_i^2E_\beta$, where $\tilde{x}_i$ depends on the matter potential. 

Note the different cross sections accompanying the second and third terms in Eq.~\eqref{eq:dNdE}, due to the different helicities of the final decay products in each case. This will produce a much larger relative enhancement of the effect of the decay in the antineutrino channel, leading to a more significant excess than in the neutrino channel.  The observation of such an effect would be extremely relevant as it would indicate that neutrinos are Majorana particles, since the opposite-helicity decay mode is not observable in the Dirac case. Also, the contributions to the number of events from visible decay are going to be weighted by the neutrino flux available at energies $E_\alpha > E_\beta$. Thus, the effect of visible decay is expected to be larger for experiments with a wide-band beam extending to high energies  as in the case of DUNE, while it will be suppressed at narrow-band beam experiments as the high-energy tail of the flux is severely suppressed by placing the detector off-axis.  

 The DUNE experiment~\cite{Acciarri:2016crz} has been simulated using using the GLoBES~\cite{Huber:2004ka,Huber:2007ji} package. The simulation details follow the Conceptual Design Report (CDR), and we have used the DUNE simulation files publicly released by the collaboration~\cite{Alion:2016uaj}. The assumed configuration uses a detector mass of 40~kton, a beam power of 1.07~MW with 80~GeV protons and a total running time of 7 years (half in neutrino and anti-neutrino modes), amounting to a total exposure of $300~\textrm{MW}\times \textrm{kton}\times \textrm{yr}$. Unless otherwise stated, the assumed true values for the oscillation parameters are:
\begin{equation}
\begin{array}{c}
\theta_{13} =  9^\circ ; \; \theta_{12} = 33^\circ ; \; \theta_{23} = 45^\circ ; \; \delta = -90^\circ  \, ,  \\[2mm]
\Delta m^2_{21} = 7.5\times 10^{-5}\textrm{eV}^2  ; \; \Delta m^2_{31} = 2.45\times 10^{-3}\textrm{eV}^2 \, ,
\end{array}
\label{eq:params}
\end{equation}
which are within the $1\sigma$ allowed regions from recent global fits to neutrino oscillation data, see \eg Refs.~\cite{Bergstrom:2015rba,Esteban:2016qun}. Since the simulations are computationally rather expensive, the analysis performed in the current work will only use the results from the appearance or disappearance channels separately, as indicated. The combination between appearance and disappearance channels is left for future studies. Normalization uncertainties have been implemented as in Ref.~\cite{Alion:2016uaj}, and minimization over the nuisance parameters has been performed. Conversely, no minimization over current uncertainties for the oscillation parameters has been included. Finally, for simplicity we will assume $|g_{31}| = |g_{32}| \equiv g $ and $|g'_{31}| = |g'_{32}| \equiv g'$ for all the results shown in this work, unless otherwise stated. 

\begin{figure}[tb!]
\begin{center}
\includegraphics[scale=0.5]{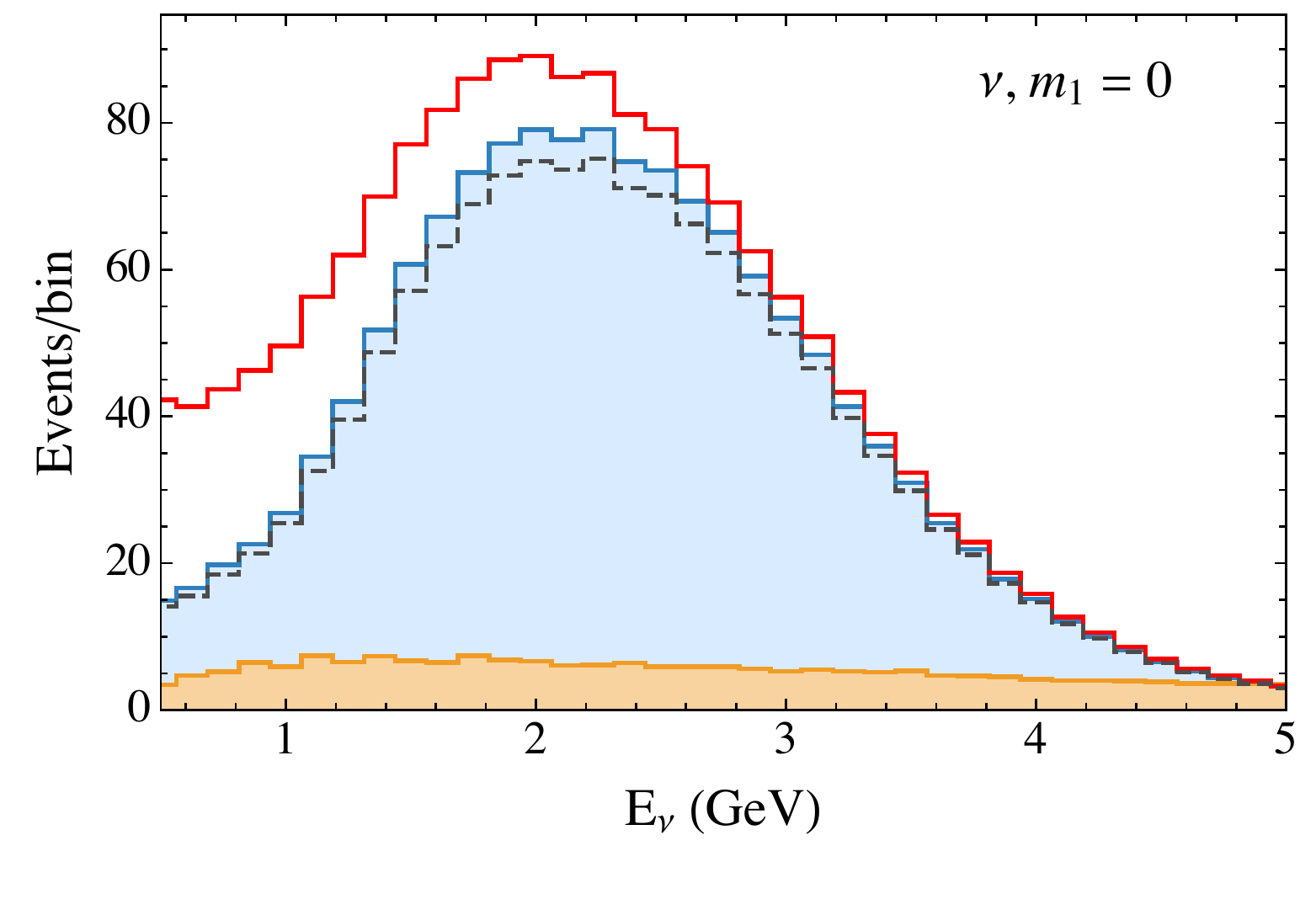}
\includegraphics[scale=0.5]{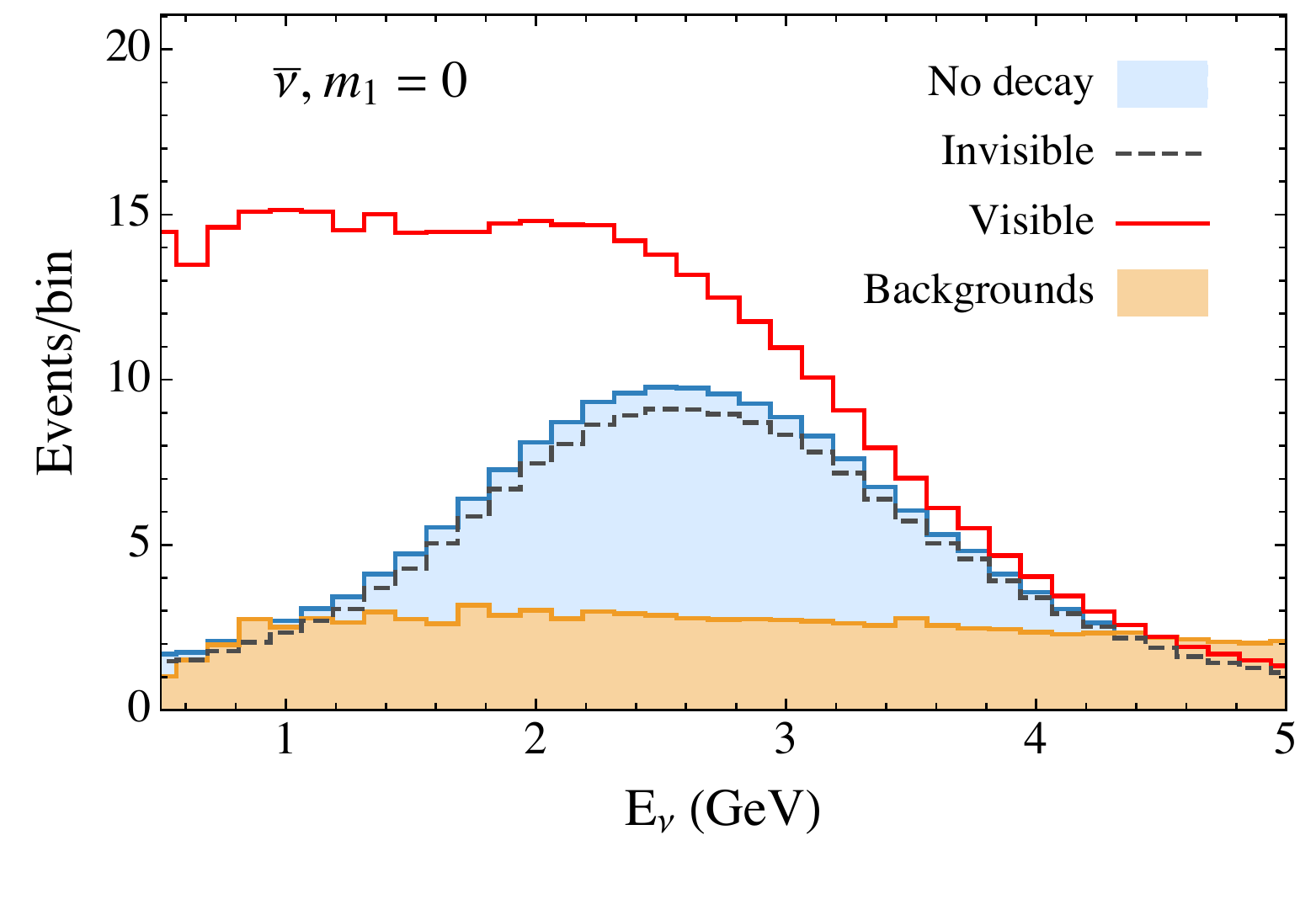}
\end{center}
\caption{\label{fig:events} Expected total event rates in the $\nu_\mu \to \nu_e$ (left) and $\bar\nu_\mu \to \bar\nu_e$ (right) oscillation channels at DUNE, as a function of the energy of the detected neutrino. Note the different scale in the two panels. The blue histograms indicate the expected signal event rates, while the dashed and solid lines show the effect of invisible and visible decay, using $g = g'= 0.5$ and a massless $\nu_1$. For comparison, the orange histogram shows the expected background event rates. The oscillation parameters used are listed in Eq.~\eqref{eq:params}, and the exposure has been fixed to $300~\textrm{MW}\times \textrm{kton}\times \textrm{yr}$. }
\end{figure}

The effect of the decay on the expected event rates can be seen in Fig.~\ref{fig:events}, for the neutrino (left panel) and antineutrino (right panel) appearance channels. The event rates are shown for DUNE for its total nominal exposure, after detection efficiencies and reconstruction effects are taken into account. In both panels, the event distributions are shown as a function of the final neutrino energy, for different scenarios: no neutrino decay (filled blue histogram), invisible decay (dashed green) and visible decay (solid red). For comparison, the expected background events are also shown by the filled orange histograms. In this figure, the mass of the lightest neutrino ($m_1$) has been set to zero and the two Majoron couplings have been set to the same value $g = g' = 0.5$. As expected, the effect from invisible decay would be a depletion in the number of events, which is largest around the oscillation maximum. Conversely, the effect of visible decay would be an excess of events at low energies, providing a very distinctive signature. The effect would be more significant in the antineutrino channel since the larger neutrino cross section produces a relative enhancement of the $\bar\nu \to \nu +\phi$ contribution, as outlined above (see Eq.~\eqref{eq:dNdE}).

\begin{figure}[t!]
\begin{center}
\includegraphics[scale=0.5]{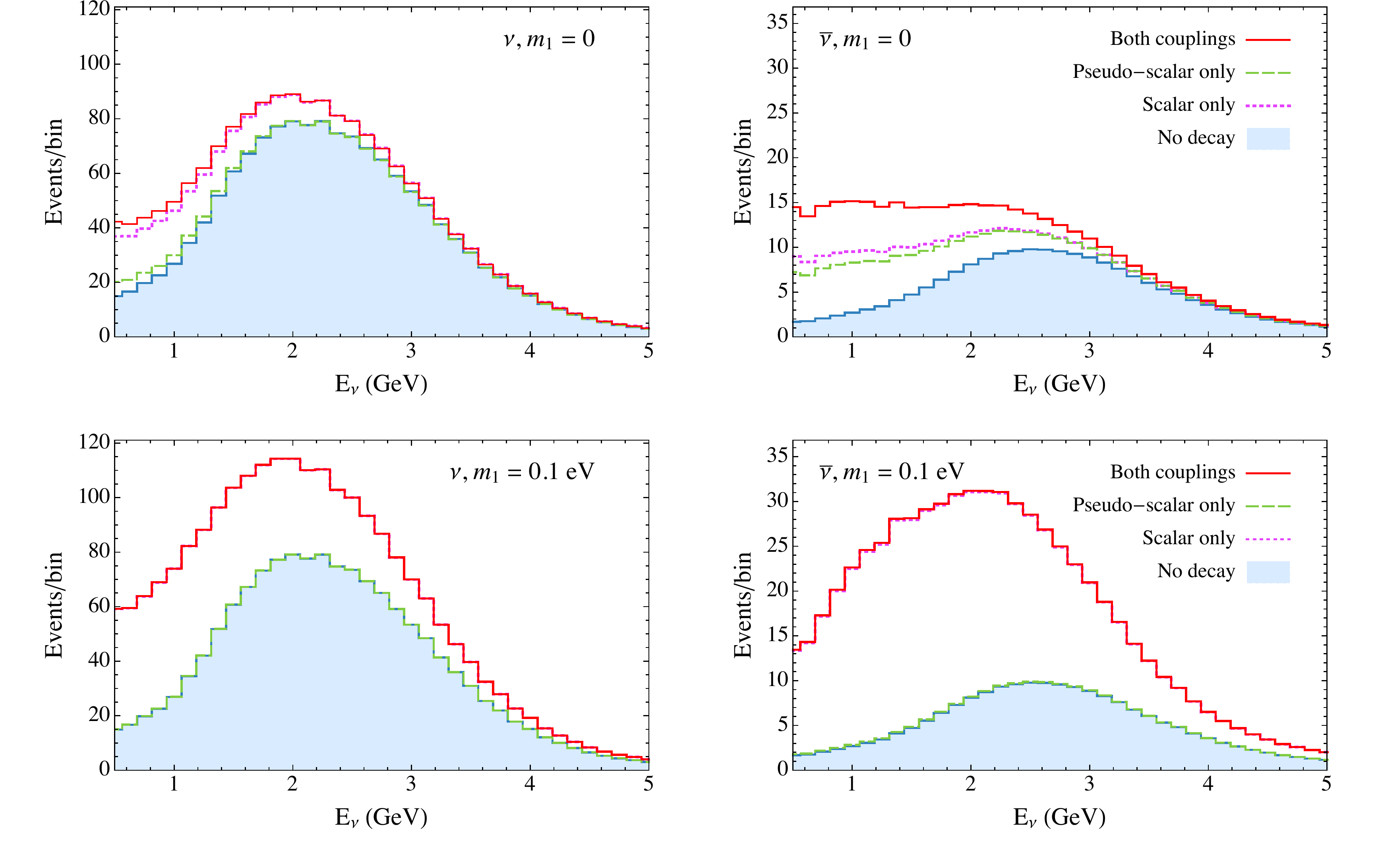}
\end{center}
\caption{\label{fig:events1} Expected total event rates in the $\nu_\mu \to \nu_e$ (left panels) and $\bar\nu_\mu \to \bar\nu_e$ (right panels) oscillation channels at DUNE, as a function of the energy of the detected neutrino. Note the different scale between left and right panels. The blue histograms indicate the expected signal event rates for stable neutrinos, while the dashed and solid lines show the effect of invisible and visible decay, using $g = 0.5$ (``Scalar only''), $ g'= 0.5$ (``Pseudo-scalar only'') and $ g= g'=0.5$ (``Both couplings''). Upper panels have been obtained assuming $ m_1 = 0$, while lower panels correspond to $ m_1 = 0.1$~eV. The oscillation parameters used are listed in Eq.~\eqref{eq:params}, and the exposure has been fixed to $300~\textrm{MW}\times \textrm{kton}\times \textrm{yr}$. }
\end{figure}

The size of the visible decay contribution to the probability, $P^{\rm vis}_{\rm mat}$, depends on the lightest mass eigenstate  $m_1$ as well as on the specific decay channel (scalar or pseudo-scalar). In Fig.~\ref{fig:events1} we show the final impact of the decay on the event distributions for different scenarios. In the upper-left (upper-right) panel we show the neutrino (anti-neutrino) event rates as a function of energy assuming that the lightest neutrino is massless. We can notice that the decay through the scalar coupling (pink curve) dominates over the pseudo-scalar one (green curves). In the lower-left (lower-right) panels we show the same distributions assuming $m_1=0.1$~eV instead. From the comparison between the upper and lower panels it can be seen that the contribution from the decay is much larger for $m_1 \neq 0$.  This can be traced back to the behaviour of $\Gamma_3 \times m_3 $ with the mass of $m_1$ as shown in Fig.~\ref{fig:gammas}. When the mass of the lightest neutrino is increased above $m_1 \sim 0.01$~eV, the decay modes with same helicity dominate the decay and the decay widths (in the lab frame) show a considerable increase with respect to the value for vanishing $m_1$. This translates into a much larger total width of the neutrino in the higher mass region than for low $m_1$ and, thus, a larger effect of the visible decay on the event rates. 

Finally, as anticipated in Sec.~\ref{sec:framework}, the scalar coupling completely dominates the decay over the pseudo-scalar coupling for $m_1 \gg 0$. This can be seen from the comparison between the dashed lines (pseudo-scalar only) and shaded histograms (no decay) in the lower panels in Fig.~\ref{fig:events1}. This  is a direct consequence of the dominance of scalar over pseudo-scalar contribution to the decay widths in the limit $x_i \to 1$ (\ie $m_1 \gg 0$), see Figs.~\ref{fig:fgk} and~\ref{fig:gammas}.  Conversely, in the limit $m_1 \to 0$ (i.e., $x_1 \to \infty$ and $x_2 \to 6$), corresponding to the upper panels in Fig.~\ref{fig:events1}, the relative contribution of the pseudo-scalar coupling is much larger, especially in the antineutrino channel where is as significant as the contribution coming from the scalar coupling in the Lagrangian. This is because of the larger neutrino cross section with respect to the antineutrino cross section, with enhances the effect of the helicity-flipping decays, as outlined above.

\section{Results}
\label{sec:results}

\begin{figure}[tb!]
\begin{center}
\includegraphics[scale=0.5]{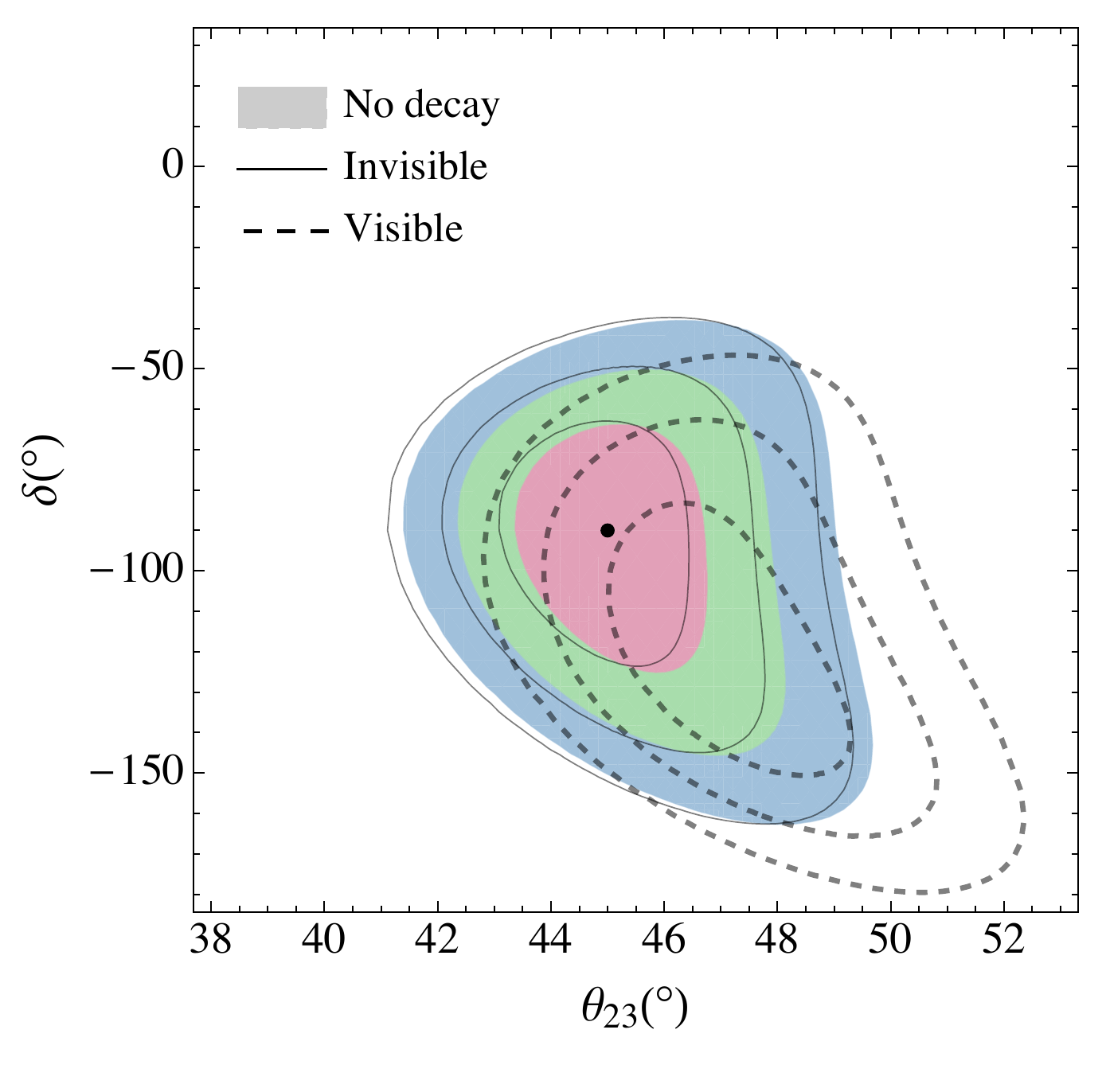}
\end{center}
\caption{ Confidence regions in the $\theta_{23} - \delta$ plane for a fit to DUNE simulated ``true'' data generated under different scenarios: no decay (colored regions), and with invisible (solid) and visible (dashed) decay, for $g = g'=0.2$ and $m_1 = 0$. In all cases, the simulated data are fitted using the standard assumption that all neutrino eigenstates are stable. The black dot indicates the assumed input values for $\theta_{23}$ and $\delta$ used to generate the simulated ``true'' data, and is the same for the three cases. The contours indicate the allowed regions at 1$\sigma$, 2$\sigma$ and 3$\sigma$ (2 d.o.f.).\label{fig:th23dcp} }
\end{figure}

First let us describe the general impact that neutrino decay would have on the fit to neutrino oscillation data, if interpreted in a three-family standard oscillation framework where the neutrinos are assumed to be stable. We show in Fig.~\ref{fig:th23dcp} the allowed confidence regions for a fit in the $\theta_{23} - \delta$  plane using appearance data alone which has been simulated under different assumptions: (i) no decay (colored regions); (ii) invisible decay (solid); (iii) and visible decay (dashed). In all cases the data are simulated using the true input values for the oscillation parameters given in Eq.~\eqref{eq:params}. In cases (ii) and (iii) the decay rates have been computed using $g =  g' = 0.2$ and $m_1 = 0.05$ eV. The resulting confidence regions from the fit differ, however, because in all cases the simulated data have been fitted using the standard assumption that all neutrino eigenstates are stable. 
%
%
As can be seen from the figure, the resulting fit is quite similar in the case where there is only invisible decay: the simulated data prefers slightly smaller values of $\theta_{23}$, which is due to the fact that the event distributions are somewhat reduced due to the effect of the decay if this is invisible, as shown in Fig.~\ref{fig:events}. The result is very different, however, if the decay products are visible. On one side, the excess of events favors larger values of $\theta_{23}$ in the fit. This comes from the pile-up of neutrino events at low energies, which can be better fitted with a larger value of the mixing angles. However, the helicity-flipping decay will produce an effect that is slightly different for neutrinos and anti-neutrinos, as explained in Sec.~\ref{sec:exp}. This creates a ``fake'' CP-violation effect in the event rates which, if interpreted under the assumption of stable neutrino mass eigenstates, can lead to a bias in the determination of the value of $\delta$. For example, in the case of visible decay the best-fit is found at $\delta = -120^\circ$ and $\theta_{23}=47^\circ$ and the fit would disfavor the true input values at approximately $1\sigma$ confidence level. Nevertheless, the minimum $\chi^2$ obtained at the best-fit point in this example is $\chi^2_{min} = 5.6$ and, thus, the experiment should be able to reject the ``no decay'' hypothesis scenario at some confidence level (for the set of couplings used in this example). 

\begin{figure}[tb!]
\begin{center}
\includegraphics[width=0.45\textwidth]{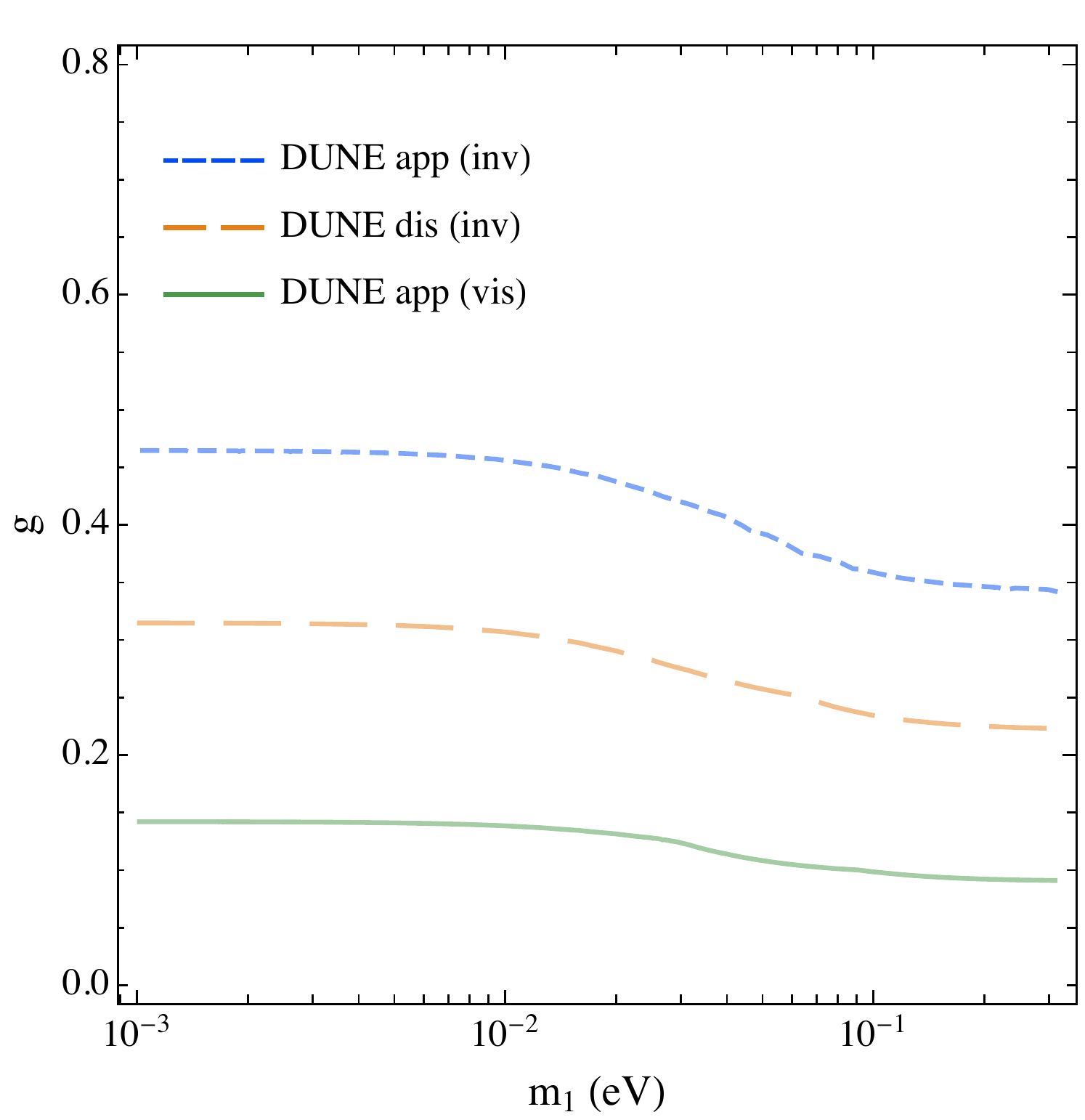}
\includegraphics[width=0.45\textwidth]{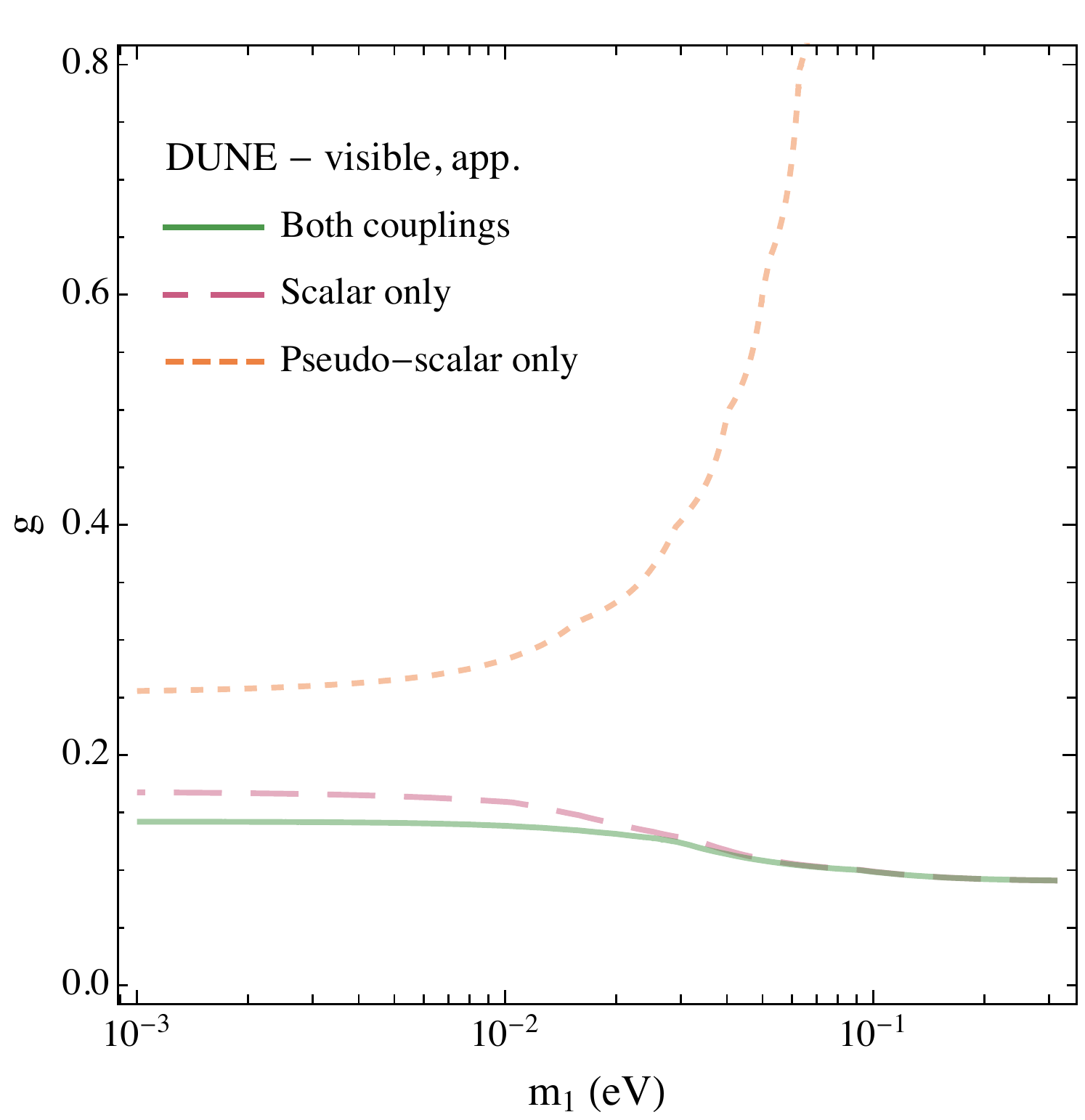}
\end{center}
\caption{\label{fig:sens-g1-g2}  DUNE sensitivities to neutrino decay using charged-current data in the appearance channels. Left panel:  sensitivities coming from appearance (app) and disappearance (dis) data separately, as indicated in the legend, for invisible (inv) or visible (vis) decay. Right panel: sensitivities assuming that only scalar ($g$) or only pseudo-scalar ($g'$) couplings are allowed at a time, as indicated in the legend. Only information coming from the appearance channels is used in this panel. The solid green line assumes that scalar and pseudo-scalar couplings take the same value ($ g = g'$), as in Fig.~\ref{fig:sens}. All limits are shown at 90\% CL for 1 d.o.f.. 
}
\end{figure}

The expected DUNE sensitivities to neutrino decay are shown in Fig.~\ref{fig:sens-g1-g2} for different cases. We have assumed that $ g = g^\prime$ for simplicity, and present our sensitivity contours in the plane $g - m_1$ at the 90\% confidence level (for 1 d.o.f.). In the left panel of  Fig.~\ref{fig:sens-g1-g2} we show different sensitivity limits expected for the DUNE setup, using: disappearance data, for invisible decay (long-dashed yellow); appearance data for invisible decay (short-dashed blue); and appearance data, for visible decay (solid green). As can be seen from the figure, in all cases there is a change in sensitivity when the mass of the lightest neutrino is increased above $m_1 \sim 0.01$~eV,  which can be traced back to the behaviour of the decay widths as a function of $m_1$, see Fig.~\ref{fig:gammas}. 
%
%
In the right panel of Fig.~\ref{fig:sens-g1-g2} we show the expected sensitivity in case only scalar (or pseudo-scalar) couplings are allowed, as indicated by the legend. As can be seen, for lower values of $m_1$ the sensitivity comes from the effect of both the scalar and pseudo-scalar couplings in the Lagrangian, while for large values of $m_1$ the sensitivity to the pseudo-scalar coupling is negligible, and the experiment is sensitive to the scalar coupling only. This can be traced back to the behaviour of the decay widths (see Fig.~\ref{fig:fgk} and Eq.~\eqref{eq:channels}), as explained in more detail in Sec.~\ref{sec:framework}.

\begin{figure}[tb!]
\begin{center}
\includegraphics[scale=0.5]{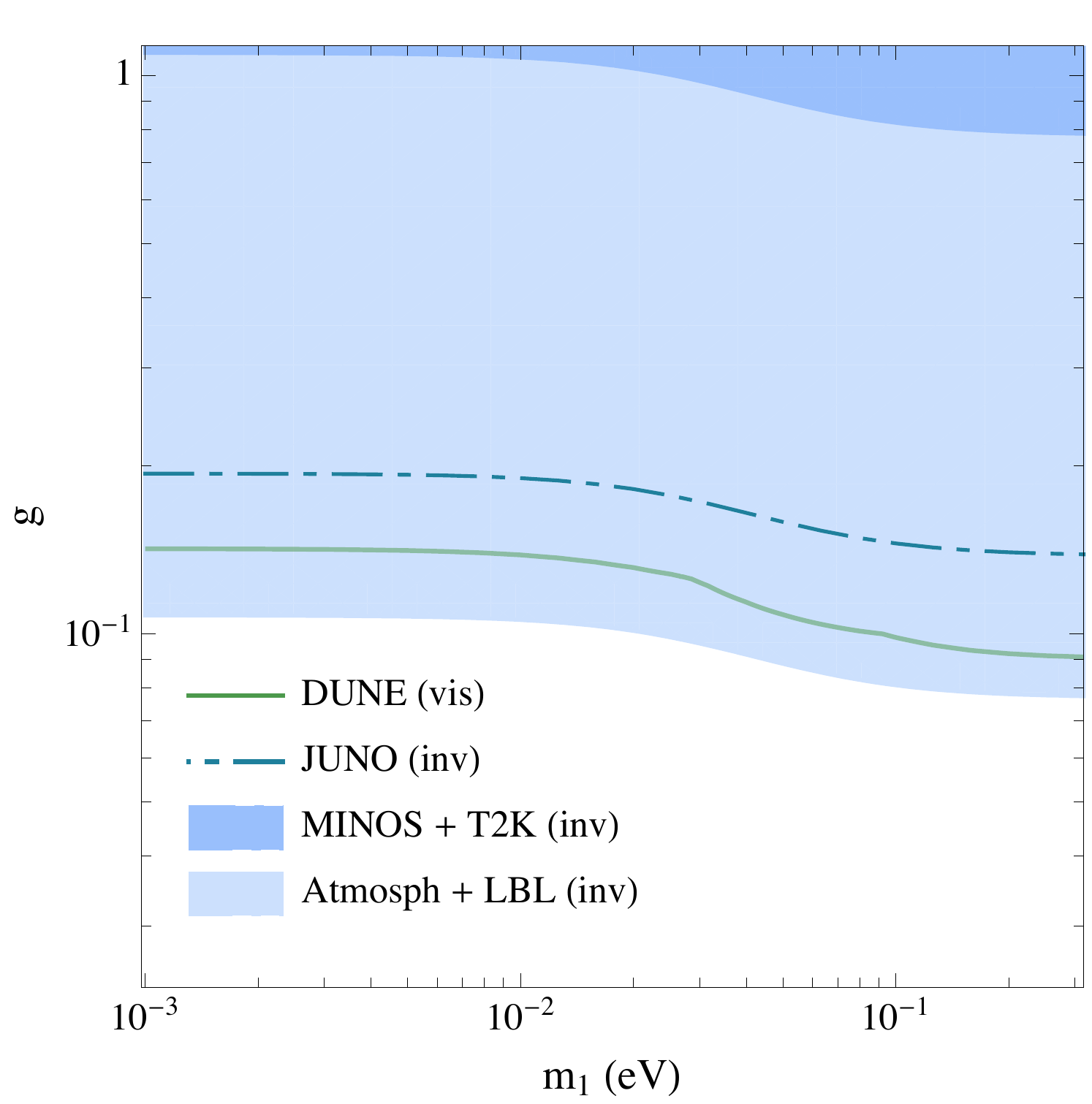}
\end{center}
\caption{\label{fig:sens} Comparison of the DUNE sensitivity to current and future limits on the lifetime of $\nu_3$. 
The dark shaded region is disfavored at the 90\% CL by MINOS CC+NC data and T2K data, $ \tau_3/m_3 > 2.8 \times 10^{-12} ~\textrm{s}/\textrm{eV}$~\cite{Gomes:2014yua}. The light shaded region is disfavored at 90\% CL from a fit to atmospheric and long-baseline data, $ \tau_3/m_3 > 2.9 \times 10^{-11}~\textrm{s}/\textrm{eV}$~\cite{GonzalezGarcia:2008ru}. Finally, for comparison we also show the expected sensitivity for the future JUNO experiment, $\tau_3/m_3 > 7.5 \times 10^{-11} ~\textrm{s}/\textrm{eV}$ (95\% C.L.), taken from Ref.~\cite{Abrahao:2015rba}. These three analyses were performed for invisible decay. The DUNE expected sensitivity (the green curve) to visible decay is computed assuming that both scalar and pseudo-scalar couplings take the same value ($ g= g'$), and using appearance data only. The sensitivity is shown at 90\% C.L. (1 d.o.f.).}
\end{figure}

Finally, Fig.~\ref{fig:sens} shows how the DUNE expected sensitivity compares to current bounds in the literature, as well as to the  expected sensitivity of the JUNO experiment. All bounds shown have been computed assuming invisible decay and therefore they constrain the combination $\tau_3/m_3 $, as discussed in Sec.~\ref{sec:framework}. To translate the bounds set on this combination into a bound in the $g-m_1$ plane we use the expression of $\Gamma_3$ as a function of the parameters $g$, $g'$ and $m_1$, following Eq.~\eqref{eq:channels}, which can be related to the value of $\tau_3/m_3$ since $\tau_3/m_3 = E/\Gamma_3$. All contours shown in Fig.~\ref{fig:sens} have been obtained assuming $g'= g$. The shaded areas have already been ruled out at 90\% C.L. by a the combination of MINOS CC and NC data and T2K CC  data~\cite{Gomes:2014yua}, or by a combination of long-baseline and atmospheric neutrino data~\cite{GonzalezGarcia:2008ru}. The future JUNO experiment will be sensitive in the region of the parameter space above the dashed blue line at 90\% C.L.. As is seen from this figure, DUNE will improve considerably with respect to the results from MINOS and T2K analysis, and will place a competitive constraint with the best direct limit currently available on the lifetime of $\nu_3$~\cite{GonzalezGarcia:2008ru}. It should be noted, however, that these results have been obtained from a fit to simulated appearance data alone.  Further improvement can be expected from the combination of appearance and disappearance data, as well as from the addition of atmospheric data. The combination of the different data samples and their impact on the sensitivity will be studied elsewhere.

\section{Summary and conclusions}
\label{sec:conclusions}

From neutrino oscillation data we know that at least two of the neutrino mass eigenstates must have nonzero masses and, therefore, can decay. Although in the Standard Model extended with neutrino masses the neutrino lifetime is longer than the age of the Universe, this may not be the case once the Standard Model is extended. In particular, since we do not know the mechanism that is responsible of neutrino masses, an interesting possibility is that it may also affect the neutrino lifetime. The observation of solar and supernova neutrinos places strong constraints on the lifetimes of the mass eigenstates $\nu_2$ and $\nu_1$. However, testing the lifetime of $\nu_3$ ($\tau_3$) is a much harder task. Currently, the strongest bounds the ratio $\tau_3/m_3$ come from a global fit to atmospheric and long-baseline data. In Ref.~\cite{Abrahao:2015rba}, the possibility of placing a constraint on the lifetime of $\nu_3$ was explored, using the future JUNO experiment as a proposed setup. In Ref.~\cite{Gago} a constraint on visible neutrino decay was placed using current MINOS and T2K data.

In most of the works mentioned above it was assumed that the neutrino decays invisibly, that is, the decay products are not observable in the detector. This would be the case, for instance, if the active neutrinos decay to lighter sterile states, or if the decay products involve active neutrinos but their energy is too low to be observable. In this work we have considered instead the possibility that the heaviest active neutrino is unstable and decays visibly. For concreteness, we have considered a model where the neutrino decays take place through a coupling with a massless scalar, usually referred to as Majoron. We have considered the most general Lagrangian, in which both scalar and pseudo-scalar couplings to the Majoron are allowed. In this case, if neutrinos are Majorana  particles, same-helicity ($\nu_i \to \nu_j +  \phi$) and opposite-helicity ($\nu_i \to \bar\nu_j + \phi$ or $\bar\nu_i \to \nu_j  + \phi$) decays would be allowed. Conversely, in the Dirac case only the decay with same helicity would be visible. This leads to an interesting phenomenology in oscillation experiments, as discussed in Sec.~\ref{sec:framework}.

\begin{table}[tbp!]
\setlength{\tabcolsep}{7pt}
\centering
\renewcommand{\arraystretch}{1.6}
\begin{tabular}{ l c c l }
\hline \hline 
Analysis & Decaying particle & Decay mode & Limit  \\
\hline 
Solar data~\cite{Picoreti:2015ika} &  $\nu_2$  & Invisible &  $\tau_2/m_2>7.2 \times 10^{-4}~\textrm{s}/\textrm{eV}$ (99\% C.L.)~ \\
Solar data~\cite{Berryman:2014qha} &  $\nu_2$  & Invisible &  $\tau_2/m_2>7.1 \times 10^{-4}~\textrm{s}/\textrm{eV}$ ($2\sigma$  C.L.)~ \\
Atmospheric and long-baseline data~\cite{GonzalezGarcia:2008ru}  & $\nu_3$ & Invisible &  $\tau_3/m_3> 2.9 \times 10^{-10}~\textrm{s}/\textrm{eV}$ (90\% C.L)\\
MINOS and T2K data~\cite{Gomes:2014yua} & $\nu_3$  &  Invisible  & $\tau_3/m_3> 2.8 \times 10^{-12}~\textrm{s}/\textrm{eV} \quad \textrm{(90\%~C.L.)}$ \\
MINOS and T2K data~\cite{Gago} & $\nu_3$  &  Visible  & $\tau_3/m_3 > 1.5 \times 10^{-11}~\textrm{s}/\textrm{eV} \quad \textrm{(90\%~C.L.)}$ \\
JUNO  expected sensitivity~\cite{Abrahao:2015rba}   & $\nu_3$ & Invisible & $\tau_3/m_3 > 7.5 \times 10^{-11}~\textrm{s}/\textrm{eV}$ (95\% C.L.) \\
DUNE expected sensitivity (this work) &  $\nu_3$ & Visible &  $\tau_3/m_3 > 1.95 - 2.6 \times 10^{-10}$s/eV (90\% C.L.) \\ \hline \hline
\end{tabular}
\caption{\label{tab:bounds} Current and prospective constrains on neutrino lifetime from neutrino oscillation experiments.}
\end{table}

The formalism for visible neutrino decay in neutrino oscillations was developed in vacuum in 
Ref.~\cite{Lindner:2001fx}, assuming that there is no significant shift in the neutrino energy spectrum for the final neutrinos. In this work, we have considered the impact of neutrino decay on the spectrum.  The observable effect on the event rates due to visible neutrino decay is a pile-up of events at low energies, as in Ref.~\cite{PalomaresRuiz:2005vf}, coming from neutrinos which were initially produced at high energies and decayed before arriving to the detector. Therefore, two experimental features are key to maximize the effect: low energy detection thresholds, and a wide-band beam extending to high neutrino energies. The DUNE experiment has both features and, therefore, a priori it can be sensitive to this type of signatures. 

Since the far detector would not be magnetized at DUNE, the number of events will contain both contributions from helicity-conserving and helicity-flipping decays (if allowed). The decay widths with opposite helicity are only sizable for $m_1 \ll 0.01$~eV, while in the limit of quasi-degenerate neutrinos this mode is suppressed. Therefore, for opposite-helicity decays the excess could be more significant in the antineutrino channel due to the enhancement of the effect via the larger neutrino cross section, as shown in Eq.~\eqref{eq:dNdE}. This offers an interesting signature which could potentially allow to distinguish its Majorana/Dirac character, if neutrino decays were observed. 

We have computed the event rates considering the effect of visible decay plus oscillations for the $\nu_\mu \to \nu_e$ and $\bar\nu_\mu \to \bar\nu_e$ oscillation channels, and fitted the results using the standard three-family scenario with stable neutrinos. For concreteness, we assumed normal ordering ($m_1 < m_2 < m_3$) and  the same size for the moduli of the scalar and pseudo-scalar couplings $g_{31}, g_{32}, g_{31}', g_{32}'$. We find that DUNE will be sensitive to visible neutrino decay for neutrino-Majoron couplings above $g > 0.14$ for a very hierarchical neutrino spectrum, and  $g> 0.09$ for quasi-degenerate neutrinos. Our bounds on the couplings imply a constraint on the lifetime of $\tau_3/m_3>1.95\times 10^{-10}$~s/eV ( $\tau_3/m_3>2.6\times 10^{-10}$~s/eV) for very hierarchical (quasi-degenerate) neutrinos. Table~\ref{tab:bounds} shows how our results compare to previous bounds in the literature as well as to prospects for the JUNO experiment. We find that DUNE will improve an order of magnitude over the constraints obtained from an analysis of charged-current T2K data and charged-current and neutral-current data from the MINOS experiment from Ref.~\cite{Gomes:2014yua}. The DUNE sensitivities are comparable to the best direct constraints on the lifetime of $\nu_3$ from oscillation experiments (taken from Ref.~\cite{GonzalezGarcia:2008ru}), as well as to the expected sensitivity of the JUNO experiment computed in Ref.~\cite{Abrahao:2015rba}. However, it should be noted that our analysis only considered charged-current data in the appearance channels. Further improvement can be  expected from the combination with neutral-current data as well as from the combination with data in the disappearance channels $\nu_\mu \to \nu_\mu$ and $\bar\nu_\mu \to \bar\nu_\mu$. DUNE will also collect a large data sample of atmospheric neutrino events, which would increase the sensitivity even further.

It should be noted that the neutrino-Majoron couplings that induce neutrino decay observable at DUNE can also be tested in meson decays~\cite{Pasquini:2015fjv} and in cosmology~\cite{Hannestad:2005ex,Bell:2005dr,Cirelli:2006kt,Friedland:2007vv,Basboll:2008fx,Archidiacono:2013dua,Farzan:2015pca,Forastieri:2015paa,Lesgourgues:2015wza,Brust:2017nmv}. Meson decays are sensitive to the Majoron couplings in the flavor basis (where cancellations can occur), while neutrino decay is in principle sensitive to the couplings in the mass basis. In cosmology, a massless scalar with $\mathcal{O}(0.1)$ couplings to neutrinos would have an impact on the number of relativistic degrees of freedom and on the neutrino free-streaming length, leading to much stronger bounds. In the present work we have not attempted to build a model capable of evading the constraints from cosmological data. Instead, we have explored the possibility of performing direct tests of neutrino-Majoron couplings using oscillation experiments, which can be complementary to other searches.

\begin{acknowledgments}
O.L.G.P. is thankful for Fermilab's hospitality and partial support under the summer visitors program.  O.L.G.P. is thankful for the support of FAPESP funding Grant  No. 2014/19164-6,  No. 2016/08308-2, FAEPEX funding grant No. 519.292, CNPQ research fellowship No. 307269/2013-2 and No. 304715/2016-6 , and for partial support from ICTP.  P.C. is thankful to the Institute  of Physics Gleb Wataghin  of Unicamp University for hospitality during her visit, where this work was initiated, and thanks FAPESP funding Grant  No. 2015/23303-4. The authors warmly thank Peter Denton, Yasaman Farzan, Gustavo Marques Tavares, Hisakazu Minakata, Hiroshi Nunokawa, Stephen Parke and Thomas Schwetz for useful discussions. Fermilab is operated by Fermi Research Alliance, LLC under Contract No. DE-AC02-07CH11359 with the United States Department of Energy. This work has received partial support from the European Union's Horizon 2020 Research and Innovation Program under the Marie Sklodowska-Curie grant agreements No. 690575 and 674896.
\end{acknowledgments}

\appendix
\section{Computation of the oscillation probabilities in presence of neutrino decay}
\label{app:decay}

\subsection*{Impact of neutrino decay on the observable neutrino spectrum} 

For the decay $ \nu_3^r \to \nu_j^s +  \phi$, the energy of the neutrino in the final state $E_\beta$ can be computed as a function of the energy of the parent neutrino $E_\alpha$. Energy conservation will render the neutrino in the final state with an energy below that of the initial neutrino, as some of its energy will be carried away by the Majoron\footnote{For simplicity, we assume that the scalar boson is massless. A massive scalar will slightly modify the decay kinematics. }. The distribution $W^{rs}_{3j}(E_\alpha,E_\beta)$  will be different depending on the decay mode:  
\beq\label{eq:W1}
W^{rs}_{3j}(E_\alpha,E_\beta) \equiv \dfrac{1}{\Gamma^{rs}_{3j}} \dfrac{d\Gamma^{rs}_{3j} (E_\alpha,E_\beta)}{dE_\beta}=
\left\{ 
\begin{array}{l l}
\left(\dfrac{1}{E_\alpha}\right) \dfrac{ g^2 (R + 2) + (g')^2 (R - 2) }{ g^2 f(x_i) + (g')^2 h(x_i) } \quad \phantom{lala}& r=s , \\[0.5cm]
\left(\dfrac{1}{E_\alpha}\right)  \dfrac{  \frac{1}{x_i} + x_i - R }{ k(x_i) } &  r \neq s \\
\end{array}
\right.
\eeq
where 
\[
  R \equiv \dfrac{1}{x_i}\dfrac{E_\alpha}{E_\beta} + x_i \frac{E_\beta}{E_\alpha}
\]
with $x_i \equiv m_3/m_i > 1$, and the functions $f(x), h(x),k(x)$ are defined in Eq.~\eqref{eq:fgk}.

In the matter regime, all relevant quantities have to be replaced by their corresponding ones computed after diagonalizing the system in presence of a matter potential. As a final remark it should be noted that, since the diagonalization renders the neutrino masses $m_2$ and $m_3$ dependent on the matter potential, their values will eventually depend on the neutrino energy. Thus, in practice, $m_3 \equiv m_3(E_\alpha)$ and $m_2 \equiv m_2 (E_\beta)$ in the expressions above.

\subsection*{Visible decay contribution to the appearance probabilities }
Here we follow mainly Refs.~\cite{Lindner:2001fx, PalomaresRuiz:2005vf}, and re-derive the oscillation probabilities considering explicitly that: (a) the energies of the parent and daughter particles do not have to coincide, and (b) there could be a sizable interference between the amplitudes for $3 \to 1$ and $3 \to 2$ decays.
For relativistic neutrinos, one can approximate:
\begin{eqnarray}
  E^{(3)} & \simeq p + \frac{m^2_3}{2E_\alpha} & \\
  E^{(1,2)} & \simeq p + \frac{m_{1,2}^2}{2E_\beta} & 
\end{eqnarray}
where we have stressed out the fact that the energy of the initial and final neutrinos does not have to (and generally will not) be the same, \ie $E_\alpha \neq E_\beta$. Within the relativistic approximation, the visible contribution to the probability reads:
\begin{eqnarray}
P_{\alpha\beta}^{\rm app} & = & \int_0^L \mid \mathcal A_{\alpha\beta} \mid^2 dL' = \nonumber \\
& = & \sum_{j = 1,2}\mid (U_{\alpha 3}^r)^* U_{\beta j}^s\mid^2 \Gamma^{rs}_{3j}  W^{rs}_{3j} \int_0^L dL 'e^{-\Gamma^{rs}_{3j} L} + \nonumber  \\
& + &  \sqrt{W^{rs}_{31}W^{rs}_{32}} \mid U_{\alpha 3}^{r} \mid^2 
2\Re  e  \left\{ \, U_{\beta 1}^s(U_{\beta 2}^s)^* \sqrt{(\Gamma_{31}^{rs})^*\Gamma_{32}^{rs}} \,\,   e^{i\tfrac{\Delta m^2_{21}L}{2 E_\beta}}  \int_0^L dL' e^{-i \tfrac{\Delta m^2_{21}L'}{2 E_\beta}}e^{-\tfrac{(\Gamma_{31}^{rs}+\Gamma_{32}^{rs})  L'}{2}} \right\} \, ,
\label{eq:P2}
\end{eqnarray}
where $ \Gamma^{rs}_{3j}$ is the decay width (in the lab frame) for $\nu_3^r \to \nu_j^s + \phi$, which depends on $E_\alpha$.
The two integrals in Eq.~\eqref{eq:P2} are straightforward to evaluate:
\begin{equation}
    \int_0^L dL' e^{-(i a + b)L'}=
\dfrac{1}{i a + b}
 \left( 1 - e^{-(i a + b)L}\right)
\label{eq:int2}
\end{equation}
where $a=0$ and $b=\Gamma_{3j}^{rs}$ for the first integral, while $a=\frac{\Delta m^2_{21}}{2 E_\beta}$ and $b =\frac{(\Gamma_{31}^{rs}+\Gamma_{32}^{rs})}{2}$ for the interference term.

After some algebra, the last part of the interference term reads: 
\begin{align}
2\Re  e \left\{U_{\beta 1}^s  (U_{\beta 2}^s)^* \sqrt{(\Gamma_{31}^{rs})^*\Gamma_{32}^{rs}}  \right. & \left. 
 e^{i\tfrac{\Delta m^2_{21}L}{2 E_\beta}} 
 \int_0^L dL' e^{-i \tfrac{\Delta m^2_{21}L'}{2 E_\beta}} e^{-\tfrac{(\Gamma_{31}^{rs}+\Gamma_{32}^{rs})  L'}{2}} \right\}  
 =  \nonumber \\
= 4 &\big | U^r_{\alpha 3} \big |^2 \big |  U^s_{\beta 1}(U^s_{\beta 2})^* \sqrt{(\Gamma^{rs}_{31})^*\Gamma^{rs}_{32}} \big | \,  
\sqrt{W^{rs}_{31}W^{rs}_{32}}   \,   \,
 \frac{ E_\beta}{ E_\beta^2 (\Gamma^{rs}_{31} +  \Gamma^{rs}_{32})^2 + (\Delta m^2_{21})^2} 
\times   \nonumber \\ 
& \times \left\{  \Delta m^2_{21} 
\left[ \sin\left( \xi + \frac{\Delta m^2_{21}L}{2 E_\beta}\right) - 
\sin\xi \, e^{-(\Gamma^{rs}_{31} + \Gamma^{rs}_{32}) L/2}\right]  
\right. + \nonumber \\ 
& \phantom{lala} \left. + 
  E_\beta (\Gamma^{rs}_{31} + \Gamma^{rs}_{32}) \left[  \cos\left( \xi + \frac{\Delta m^2_{21}L}{2 E_\beta}\right) - 
  \cos \xi \, e^{-(\Gamma^{rs}_{31} + \Gamma^{rs}_{32}) L/2} \right]   \right\} \, .
\label{eq:interf2}
\end{align}
where we have defined $\xi = \textrm{arg}(U_{\beta 1}^s (U_{\beta 2}^s)^* \sqrt{(\Gamma^{rs}_{31})^*\Gamma^{rs}_{32}})$.
Substituting Eqs.~\eqref{eq:int2} and~\eqref{eq:interf2} into Eq.~\eqref{eq:P2}, we finally get the expression in Eq.~\eqref{eq:Pvis}.

In presence of matter effects, the expressions are obtained in the same way just replacing the decay widths and eigenvalues by the corresponding ones in matter ($U^{r}\to \tilde{U}^{r}$, $\Delta m^2_{ij} \to \Delta \tilde{m}^2_{ij}$, $\Gamma_3 \to \tilde{\Gamma}_3$). In this case the diagonalization is no longer done via a unitary matrix and, therefore, $(\tilde{U}^{r})^{-1} \neq (\tilde{U}^{r})^{\dagger}$. A final note is needed regarding the extraction of the eigenvalues in presence of neutrino decay in the matter regime.  As we have argued, the diagonalization of the Hamiltonian leads to a matrix that is not unitary. It will also lead to eigenvalues which deviate from the ones obtained in matter with no decay (see \eg Fig.~1 in Ref.~\cite{Denton:2016wmg}).


\bibliographystyle{JHEP}
\bibliography{draft-decay}
\end{document}